\documentclass[a4paper,fleqn,usenatbib]{mnras}
\usepackage{newtxtext,newtxmath}
\usepackage[T1]{fontenc}
\usepackage{ae,aecompl}
\usepackage{graphicx}	
\usepackage{amsmath}	
\usepackage{amssymb}	
\usepackage[utf8]{inputenc}
\usepackage{amsfonts}
\usepackage{mathrsfs}
\usepackage{amsmath}
\usepackage{graphicx}	
\usepackage{amssymb}	
\usepackage{subfigure}
\usepackage{caption}
\usepackage{dsfont}
\usepackage{bm}
\usepackage[usenames, dvipsnames]{color}

\title[Anisotropic clustering in the SDSS-IV eBOSS DR14 quasar sample]{The clustering of the SDSS-IV extended Baryon Oscillation
Spectroscopic Survey DR14 quasar sample: anisotropic clustering analysis in configuration-space}
\author[J. Hou et al.]{
Jiamin Hou$^{1,2}$\thanks{E-mail: jiamin.hou@mpe.mpg.de},
Ariel G. S\'anchez$^{2}$, Rom\'an Scoccimarro$^{3}$, Salvador Salazar-Albornoz$^{1,2}$,
\newauthor Etienne Burtin$^{4}$, H\'ector Gil-Mar\'in$^{5,6}$, Will J. Percival$^{7}$, Rossana Ruggeri$^{7}$, 
\newauthor Pauline Zarrouk$^{4}$, Gong-Bo Zhao$^{7,8,14}$, Julian Bautista$^{7}$, Jonathan Brinkmann$^{13}$, 
\newauthor Joel R. Brownstein$^{12}$, Kyle S. Dawson$^{12}$, N. Chandrachani Devi$^{18}$,
Adam~D.~Myers$^{9}$, 
\newauthor Salman Habib$^{19}$, Katrin Heitmann$^{19}$, Rita Tojeiro$^{15}$, Graziano Rossi$^{16}$,  
\newauthor Donald P. Schneider$^{10,11}$, Hee-Jong Seo$^{17}$ and Yuting Wang$^{8}$
\\
$^{1}$Universit\"{a}ts-Sternwarte M\"{u}nchen, Ludwig-Maximilians-Universit\"{a}t Munchen, Scheinerstraße 1, 81679 M\"{u}nchen, Germany\\
$^{2}$Max-Planck-Institut für extraterrestische Physik, Postfach 1312, Giessenbachstr., 85741 Garching, Germany\\
$^{3}$Center for Cosmology and Particle Physics Department of Physics New York University, NY 10003, New York, USA\\
$^4$ IRFU,CEA, Universit´e Paris-Saclay, F-91191 Gif-sur-Yvette, France\\
$^5$ Sorbonne Universit\'es, Institut Lagrange de Paris (ILP), 98 bis Boulevard Arago, 75014 Paris, France \\
$^6$ Laboratoire de Physique Nucléaire et de Hautes Energies, Universit\'e Pierre et Marie Curie, 4 Place Jussieu, 75005 Paris, France \\
$^{7}$Institute of Cosmology $\&$ Gravitation, Dennis Sciama Building, University of Portsmouth, Portsmouth, PO1 3FX, UK\\
$^{8}$National Astronomy Observatories, Chinese Academy of Science, Beijing, 100012, P.R. China\\
$^{9}$ Department of Physics and Astronomy, University of Wyoming, Laramie, WY 82071, USA\\
$^{10}$  Department of Astronomy and Astrophysics, The Pennsylvania State University, University Park, PA 16802\\
$^{11}$ Institute for Gravitation and the Cosmos, The Pennsylvania State University, University Park, PA 16802\\
$^{12}$ Department of Physics and Astronomy, University of Utah, 115 S. 1400 E., Salt Lake City, UT 84112, USA\\
$^{13}$  Apache Point Observatory, P.O. Box 59, Sunspot, NM 88349\\
$^{14}$ College of Astronomy and Space Sciences, University of Chinese Academy of Sciences, Beijing 100049, China \\
$^{15}$ School of Physics and Astronomy, University of St Andrews, North Haugh, St Andrews KY16 9SS, UK\\
$^{16}$ Department of Physics and Astronomy, Sejong University, Seoul, 143-747, Korea\\
$^{17}$ Department of Physics and Astronomy, Ohio University, Clippinger Labs, Athens, OH 45701\\
$^{18}$ Instituto de Astronom\'{i}a, Universidad Nacional Aut\'{o}noma de M\'{e}xico, Box 70-264, M\'{e}xico City, M\'{e}xico\\
$^{19}$ HEP and MCS Divisions, Argonne National Laboratory, Lemont, IL 60439, USA\\
}

\date{Accepted XXX. Received YYY; in original form ZZZ}

\pubyear{2018}

\begin{document}
\label{firstpage}
\pagerange{\pageref{firstpage}--\pageref{lastpage}}
\maketitle

\begin{abstract}
   {We explore the cosmological implications of anisotropic clustering measurements of the quasar sample from Data Release 14 of the Sloan Digital Sky Survey IV Extended Baryon Oscillation Spectroscopic Survey (eBOSS) in configuration space. The $\sim 147,000$ quasar sample observed by eBOSS offers a direct tracer of the density field and bridges the gap of previous BAO measurements between redshift $0.8<z<2.2$. }
   {By analysing the two-point correlation function characterized by clustering wedges $\xi_{\rm w_i}(s)$ and multipoles $\xi_{\ell}(s)$, we measure the angular diameter distance, Hubble parameter and cosmic structure growth rate. }
  {We define a systematic error budget for our measurements based on the analysis of 
  $N$-body simulations and mock catalogues.}
  {Based on the DR14 large scale structure quasar sample at the effective redshift 
$z_{\rm eff}=1.52$, we find the growth rate of cosmic structure 
$f\sigma_8(z_{\rm eff})=0.396\pm 0.079$, and the 
geometric parameters $D_{\rm V}(z)/r_{\rm d}=26.47\pm 1.23$, and $F_{\rm AP}(z)=2.53\pm 0.22$, 
where the uncertainties include both statistical and systematic errors.
These values are in excellent agreement with the best-fitting standard ${\rm \Lambda CDM}$ model 
to the latest cosmic microwave background data from Planck.
}   
\end{abstract}


\section{Introduction}
\label{sec: Introduction}

In the standard cosmological picture, the baryonic material in the early universe 
forms a hot plasma as it is tightly coupled to the photons via Compton scattering. 
Primordial inhomogeneities produce spherical acoustic waves that propagate outward from 
overdense regions. As the universe expands, the photons and baryonic matter decouple at 
the epoch of recombination and freeze the acoustic waves \citep{Peebles-1970, Sunyaev-1970}, 
which leave an imprint on the large-scale structure (LSS) of the Universe known as the 
baryon acoustic oscillations (BAO). This feature can be detected by analysing two-point 
statistics of the matter distribution, such as the power spectrum or the correlation function
\citep{Eisenstein1998,Meiksin1999,Matsubara2004}. 
Since the scale associated with the BAO feature is closely related to the 
sound horizon at the drag redshift, $r_{\rm d}\simeq150\,{\rm Mpc}$, it can be used as a robust standard ruler 
to measure cosmic distances. 
Measurements of the BAO scale in the directions parallel and perpendicular to the line of 
sight at different redsfhits can be used to probe the redshift evolution of the 
Hubble parameter, $H(z)$, and the angular diameter distance, $D_{\rm M}(z)$,
through the Alcock--Paczynski (AP) test \citep{Alcock-1979,Blake-2003,Linder-2003}.

At low and intermediate redshifts, $z\lesssim 1$, BAO measurements can be obtained using 
galaxies as tracers of the LSS of the Universe.
The first detections of the BAO signal in LSS by \citet{Cole-2005} and \citet{Eisenstein-2005}, used data from the Two-degree Field Galaxy Redshift survey \citep[2dFGRS,][]{Colless2001,Colless2003}
and the luminous red galaxy sample of the Sloan Digital Sky Survey \citep[SDSS,][]{York2000}, 
respectively. Present-day distance measurements based on galaxy clustering have reached 
per-cent level precision \citep{Anderson2012, Anderson2013, Anderson2014, Alam2017}. At higher redshift, $z\sim 2.5$, the auto-correlation of HI 
absorption lines \citep{Busca-2013, Delubac-2015, Bautista-2017} and cross-correlation with 
quasars \citep{Font-Ribera-2014} have also been used to detect the BAO signal.

Clustering measurements based on galaxy redshift surveys provide additional cosmological 
information beyond that contained in the BAO feature. A particularly important source
of information is the signature of the so-called redshift-space distortions (RSD), 
induced by the line-of-sight component of the peculiar velocities of the galaxies.
As the peculiar velocity field is sourced by the matter overdensity, the analysis 
of the resulting pattern of anisotropies in the clustering of the tracers can be used to 
constrain the growth rate of cosmic structures, usually expressed in terms of 
the combination $f\sigma_8(z)$ \citep{Guzzo2008}.

In this work, we employ quasars as tracers of the LSS of the Universe.
Quasars, whose luminosities are powered by supermassive black holes at their centres, 
are intrinsically much more luminous than galaxies and can be detected at higher redshifts. 
Thus, they open a new redshift range for LSS clustering analyses. 
The Data Release 14 (DR14) quasar sample from the extended Baryon Oscillation Spectroscopic 
survey \citep[eBOSS,][]{Dawson-2016}, covers the redshift range $0.8<z<2.2$, 
bridging the gap between the measurements inferred from the clustering of galaxies 
and those recovered from the Ly-$alpha$ forest of high redshift quasars. 
We characterize the spatial distribution of the eBOSS DR14 quasar sample by means of 
clustering statistics in configuration space. We measure the two-point correlation function 
and decompose it into 
Legendre polynomials \citep{Padmanabhan-2008, Samushia-2014} and clustering wedges 
\citep{Kazin-2013, Sanchez-2013, Sanchez-2014, Sanchez-2017a}. The analysis of the full shape 
of these measurements allow us to exploit the joint information from BAO and RSD, which 
we compress into measurements of the geometric parameter combinations 
$D_{\rm V}(z)/r_{\rm d}$, where $D_{\rm V}(z)\propto \left(D_{\rm M}(z)^2/H(z)\right)^{1/3}$,
and $F_{\rm AP}\propto D_{\rm M}(z)H(z)$, and the growth rate parameter $f\sigma_8(z)$.

This work is part of a series of papers analysing the anisotropic clustering pattern of the 
DR14 LSS quasar sample \citep{Gil-Marin-2017, Ruggeri-2017, Zhao-2017, Zarrouk-2017}.
Of these analyses, those of \citet{Gil-Marin-2017} and \citet{Zarrouk-2017}
are more similar to this paper. 
\citet{Gil-Marin-2017} use the RSD model of \citet{Taruya-2010} to extract cosmological 
information from the full shape of Legendre multipoles in Fourier space. \citet{Zarrouk-2017} 
use configuration-space clustering measurements identical to the ones in this paper, 
but applying a different model based on convolution Lagrangian perturbation theory 
\citep[CLPT,][]{Carlson-2013} and the Gaussian streaming model of RSD \citep{Reid-2011}. 
The analyses of \cite{Ruggeri-2017} and \cite{Zhao-2017} are based on Fourier-space 
measurements, but computed after applying a set of redshift-dependent weights to the QSO eBOSS 
catalogues that allow for lossless compression of the information along the redshift direction.  A full comparison between the conventional analyses and the redshift-weighted methods can be found in \citet{Zarrouk-2017}.

This paper is structured as follows: Section \ref{sec: data} provides detailed information on 
the eBOSS DR14 quasar survey, presents our anisotropic clustering measurements based on 
this sample, and describes our methodology to obtain cosmological constraints out of them. 
This section also introduces the mock catalogues that are used to 
estimate the covariance matrix of our clustering measurements and for model testing. 
Section \ref{sec: the_model} contains a short review of our model of the anisotropic 
correlation function and its validation using the mock catalogues and $N$-body simulations.
In Section \ref{sec: constraints} we  presents the geometric 
constraints and measurements of the growth of structure derived from the eBOSS quasar sample. Finally, Section~\ref{sec: conclusion} presents a summary of our main results and our 
conclusions.

\section{The clustering of quasars in eBOSS}
\label{sec: data}

\subsection{The extended Baryon Oscillation spectroscopic survey}
\label{sec: survey}

The eBOSS survey is a part of SDSS-IV \citep{Blanton-2017} and mainly focuses 
on mapping the distribution of large scale structure using a variety of tracers: luminous red galaxies (LRGs) $0.6 < z < 0.8$, emission line 
galaxies (ELGs) $0.7 < z < 1.1$ and a low-redshift quasar sample at $0.8<z<2.2$ (hereafter, 
LSS quasars) that is the focus of this paper. The eBOSS DR14 quasar sample consists of two sky regions, with $116,866$ objects in the Northern Galactic Cap (NGC) and $77,935$ in the 
Southern Galactic Cap (SGC). The survey footprint covers an area of ca. $2112.92 \,{\rm deg}^2$ with a mean completeness of $\sim 0.97$.

The eBOSS quasar candidates are selected through the imaging data from SDSS-I/II/III \citep{Gunn-2006}, the $2.5$-meter Sloan Telescope and the Wide Field Infrared Survey Explorer
\citep[WISE,][]{Wright-2010}. In order to facilitate clustering measurements, the sample is selected homogeneously with a comoving number density of $\bar{n}\simeq 10^{-5} ({\rm Mpc}/h)^3$. The ``CORE'' selection is performed by a likelihood-based routine {\it 
extreme deconvolution} \citep[XDQSOz,][]{Bovy-2012} over five broad bands {\it ugriz}, with a 
mid-IR-optical color cut from WISE imaging to help distinguish quasars from stars\citep{Myers-2015}. Apart from the ``CORE'' sample, another selection based on variability in multi-epoch imaging from the Palomar Transient Factory  \citep[PTF][]{Rau-2009} was also applied.

The targets are observed by the BOSS double-armed spectrographs \citep{Smee-2013}. The DR14 LSS quasar catalogue comes from three sources: 1. Legacy survey, a previous SDSS project, with confident redshift measurements. 2. SEQUELS, a pilot survey for eBOSS started during SDSS-III, 3. eBOSS, that contains over 75 $\%$ of the redshifts in the DR14 LSS catalogues.

The selected quasar targets and the corresponding redshift information are combined to construct the LSS quasar catalogue. {The redshift estimate starts with the SDSS pipeline, which is based on principal component analysis. When the identification and redshift of a target is considered inaccurate, a further visual inspection is applied. If the ${\rm MgII}$ emission line is present at a spectra, its peak is used as an estimator of the redshift. The redshift estimate based this broad emission line is considered as the most robust estimate given the redshift range of the DR14 sample. Otherwise, the peak of ${\rm CIV}$ is used \citep{Paris-2012} but this line is potentially affected by the quasar outflow \citep{Hewett-2010, Shen-2016}. The uncertainty in redshift determination can have an impact on the clustering measurement, given that our sample sits at a relatively hight redshift. This effect needs to be taken into account in our clustering modelling and we will further stress this point in sec. \ref{subsec: model of CorrFunc}.}

\subsection{Anisotropic clustering measurements}
\label{subsec: wedges and multipoles}

\begin{figure*}
\centering
\includegraphics[width=\columnwidth]{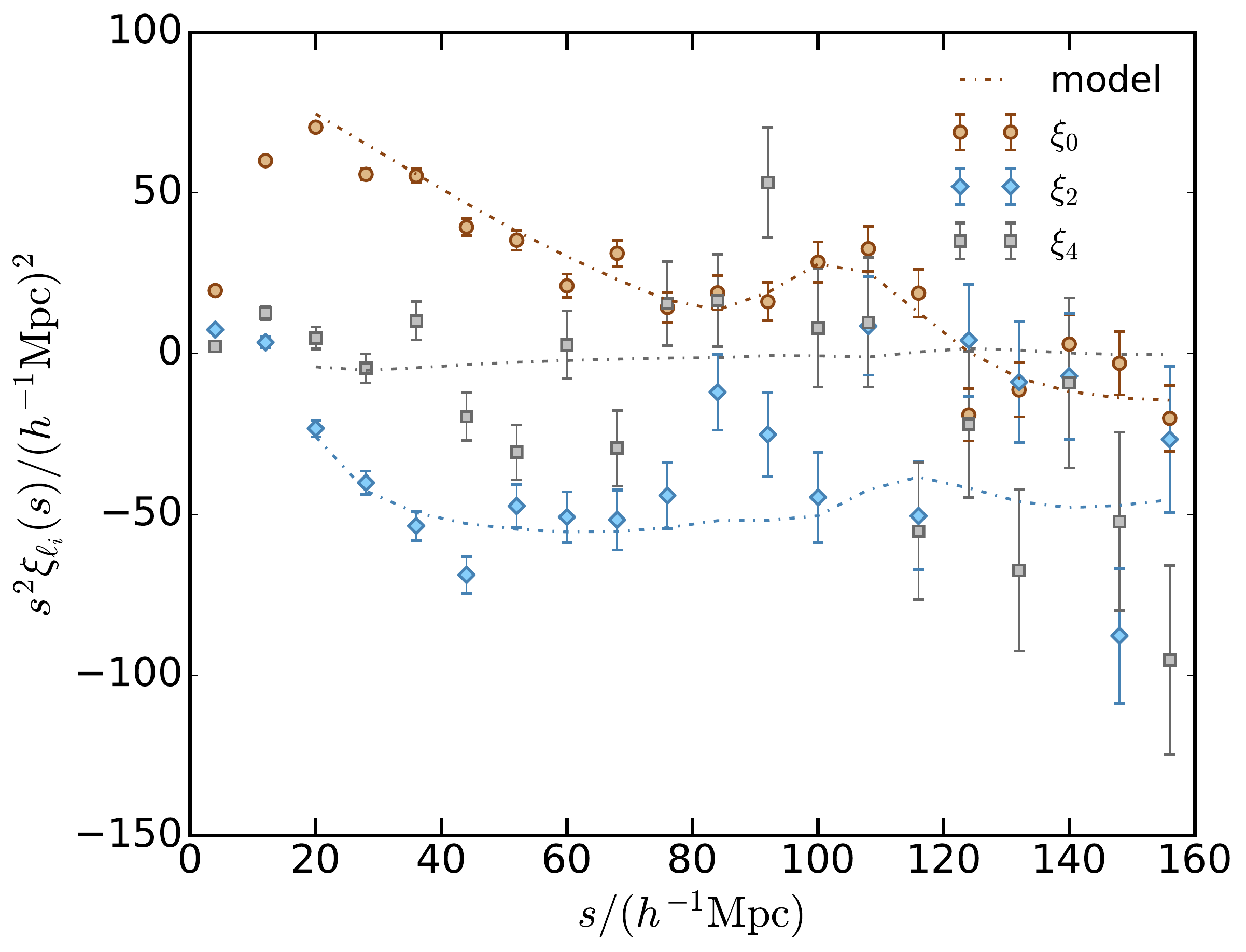}
\includegraphics[width=\columnwidth]{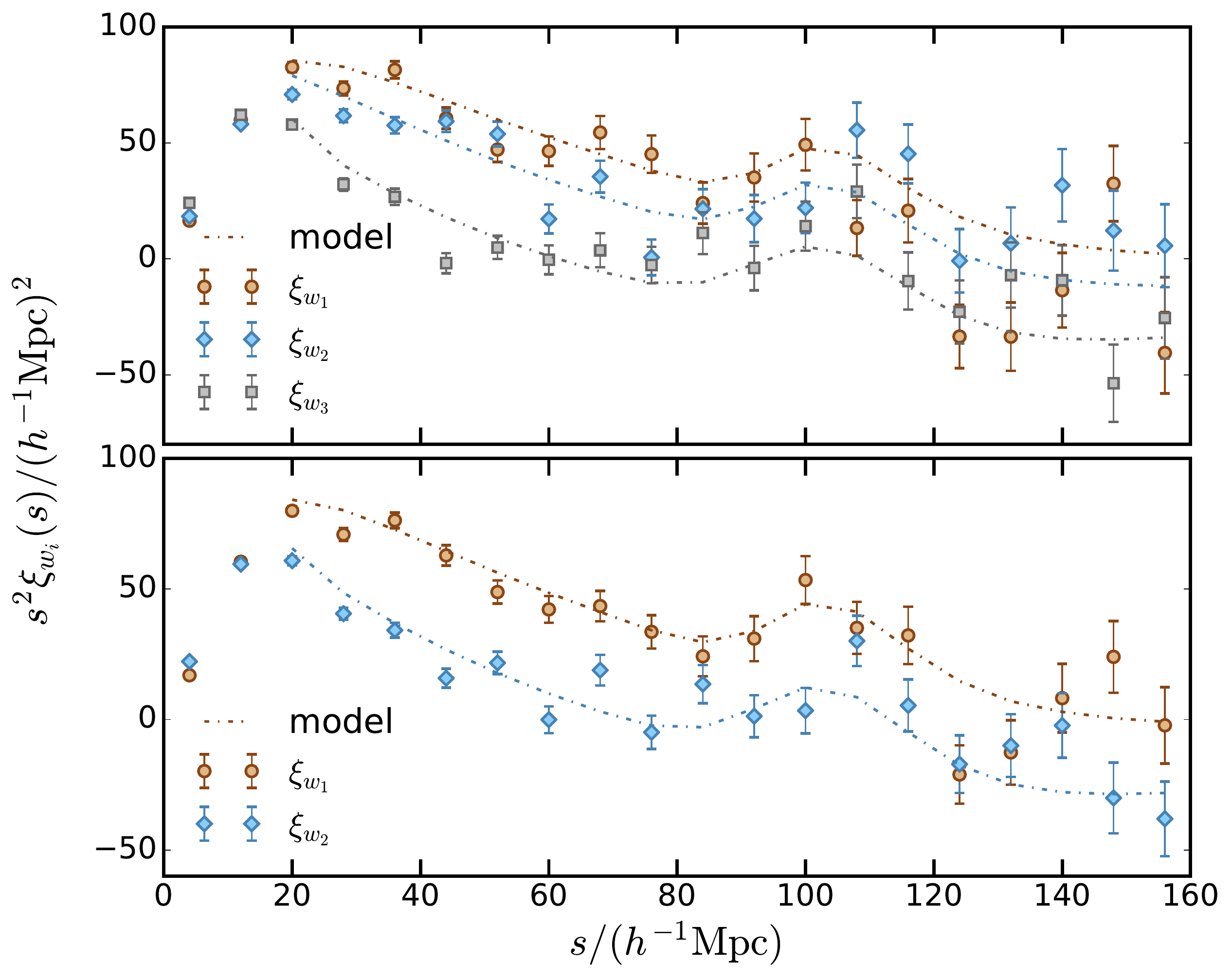}
\caption{Left: Legendre multipoles $N_{\ell_i}=3$, monopole(red) $\ell=0$, quadrupole (cyan) $\ell=2$, and hexadecapole (grey) $\ell=4$. Right: upper panel displays clustering wedges $N_{w_i}=3$ in the directions parallel (red) intermediate (cyan) and transverse (grey) to the line of sight and lower panel shows $N_{w_i}=2$ without the intermediate wedge. The multipoles and wedges are measured from the quasar sample of eBOSS DR14. The dashed lines correspond to the best fitting model to these measurements. The errorbars are inferred from $10^3$ sets of mock catalogues (EZmocks).}
\label{fig: obs_2ptcf_ds8}	   
\end{figure*}

The correlation function $\xi(\mathbf{s})$ characterises the probability (in excess of random) 
of observing pairs of galaxies as a function of their separation, $\bf{s}$. 
Assuming rotational symmetry along the line of sight direction, the correlation function is 
reduced to the two-dimensional function $\xi(\mathbf{s})\equiv \xi(\mu,s)$, where $\mu=\cos(\theta)$, and $\theta$ is the angle between the separation vector $\mathbf{s}$ and the line of sight direction. The analysis of the full two-dimensional correlation function $\xi(\mu,s)$ poses two problems: its low signal-to-noise ratio and the large size of its covariance matrix. Fortunately, the information of the full anisotropic correlation function can be condensed into a small set of one-dimensional projections, such as the Legendre multipoles obtained by expanding $\xi(\mu,s)$ in terms of Legendre polynomials, given by
\begin{equation}
\xi_{\ell}(s)\equiv \frac{2\ell+1}{2} \int^{1}_{-1} \xi(\mu,s) L_{\ell}(\mu) \,{\rm d}\mu,
\label{eqn: multi_def}
\end{equation}
or alternatively, by computing angular averages over wide $\mu$-bins, commonly referred to 
as clustering wedges \citep*{Kazin-2012}, 
\begin{equation}
\xi_{\Delta\mu}(s)\equiv \frac{1}{\Delta\mu} \int^{\mu_{\rm 2}}_{\mu_{\rm 1}} \xi(\mu,s) \,{\rm d}\mu,
\label{eqn: wedge_def}
\end{equation}
where $\Delta\mu=\mu_2-\mu_1$. These statistics are related by 
\begin{equation}
 \label{eqn: xi_w_from_ell}
 \xi_{\Delta\mu}(s) = \sum_\ell \xi_\ell(s) \, {\bar L}_\ell,
\end{equation}
where ${\bar L}_\ell$ is the average of the Legendre polynomial of order $\ell$ over the $\mu$-bin of the clustering wedge. 
We consider measurements of the Legendre monopole, quadrupole and hexadecapole moments 
($\ell = 0$, 2 and 4), as well as of wedges defined in terms of two and three wide angular bins obtained by dividing the $\mu$ range from 0 to 1 into two and three equal-width intervals. We refer to the individual wedges obtained in this manner by $\xi_{n{\rm w},i}(s)$, with $n=2, 3$, for the intervals $(i-1)/n < \mu < i/n$.

Note that, the observed quasar density in the eBOSS catalogue is affected by the systematic effects, observing and targeting strategies. Therefore, a series of weights need to be applied to correct for these effects,

\begin{itemize}
\item Systematic weight $w_{\rm sys}$ is introduced to remove the Galactic extinction and magnitude limiting dependency.
\item Close pair weight $w_{\rm cp}$ is used to upweight a quasar in case the projected spatial separation between this quasar and its close neighbour is below the fiber resolution
\item Focal plane weight $w_{\rm fc}$ corrects for the failure in obtaining redshift due to the position of the fiber with respect to the focal plane coordinate.
\item A radial weight $w_{\rm FKP}$ \citep{Feldman-1994} is applied to minimise the variance of measurement, $w_{\rm FKP}=\left({1+P_0 n(z)}\right)^{-1}$, where we have set $P_0=6000 h^{-3}{\rm Mpc}^3$ and $n(z)$ is the expected number density as a function of redshift.
\end{itemize} 
The final weight applied to the objects is defined by, 
\begin{equation}
w_{\rm tot}=w_{\rm FKP}\cdot w_{\rm sys}\cdot w_{\rm cp}\cdot w_{\rm fp}. 
\label{eqn: weight_tot}
\end{equation}

Fig. \ref{fig: obs_2ptcf_ds8} shows the Legendre multipoles $\xi_{\ell=0,2,4}(s)$ (left panel) and clustering wedges (right panel) as a function of the pair separation with binning of 
$ds=8\,h^{-1}{\rm Mpc}$. 
The error bars correspond to the square root of the diagonal elements of the 
covariance matrices of these measurements, computed as described in 
Section~\ref{subsec: covmat and EZ}.
The BAO signal can be observed as a bump at scale ${\rm ds}\sim 110 h^{-1}{\rm Mpc}$ both for the monopole on the left panel and the $\mu$-wedges on the right panel. The dashed line in the figure corresponds to the best fit to the data points using the theoretical model described in sec. \ref{subsec: model of CorrFunc}.

We first measured the full two-dimensional correlation function $\xi(\mu,s)$ of the quasar sample using the estimator of \citet{Landy-1993} and computed the Legendre multipoles and $\mu$-wedges using equations (\ref{eqn: multi_def}) and (\ref{eqn: wedge_def}). We employed a random catalogue following the same selection function as the real eBOSS data, but containing 40 times more objects. 

The redshift of each quasar in the catalogue was transformed into comoving distances by assuming a fiducial cosmology. 
In agreement with the fiducial cosmology used in \citep{Ata-2017}, we assume a flat $\Lambda$CDM model with matter density $\Omega_{\rm m}=0.31$, baryon density $\Omega_{\rm b}h^2=0.022$, total neutrino mass $\sum m_{\nu}=0.06 \rm eV$, and a Hubble parameter $h=0.676$. 

Any difference between the true and fiducial cosmologies leads to a rescaling of the 
components of the separation vector ${\bf s}$ in the direction transverse and parallel to the line 
of sight, $s_\perp$ and $s_{\parallel}$, by the geometric distortion factors $q_{\perp}$ and $q_{\parallel}$, given by 
\begin{align}
 q_{\perp} &= \frac{D_{\rm M}(z_{\rm m})}{D'_{\rm M}(z_{\rm m})}, \label{eq:q_perp}\\
 q_{\parallel} &= \frac{H'(z_{\rm m})}{H(z_{\rm m})}.\label{eq:q_para}
\end{align}
This rescaling distorts the shape of the measured correlation function $\xi(s,\mu)\rightarrow
\xi(s^\prime,\mu^\prime)$, with \citep{Ballinger-1996}
\begin{equation}
s=s^{\prime}\sqrt{q^2_{\parallel}(\mu^{\prime})^2+q_{\perp}(1-{\mu^{\prime}}^{2})},
\label{eqn: fid_s}
\end{equation}
and
\begin{equation}
\mu=\frac{q_{\parallel}\mu^{\prime}}{\sqrt{q^2_{\parallel}(\mu^{\prime})^2+q^2_{\perp}
(1-{\mu^{\prime}}^2)}}.
\label{eqn: fid_mu}
\end{equation}
The geometric distortions described by equations (\ref{eqn: fid_s}) and (\ref{eqn: fid_mu}) 
are the basis of the use of the BAO signal in the directions transverse and parallel to 
the line of sight to obtain measurements of the angular diameter distance $D_{\rm M}(z_{\rm m})$ 
and the Hubble parameter $H(z_{\rm m})$ \citep{Hu-2003,Blake-2003,Linder-2003}.
As the intrinsic BAO position depends on the sound horizon at drag epoch, $r_{\rm d}$, 
the information the measurements is often expressed in terms of 
rescaling parameters that include the fiducial sound horizon,
\begin{equation}
\alpha_{\perp} = q_{\perp}\frac{r_{\rm d}^{\rm fid}}{r_{\rm d}}\qquad {\rm and}\qquad
\alpha_{\parallel} = q_{\parallel}\frac{r_{\rm d}^{\rm fid}}{r_{\rm d}},
\label{eqn: AP_params}
\end{equation}
which are commonly referred to as as Alcock-Paczynski parameters (AP) \citep{Alcock-1979}.
It is worth noticing that when fitting the full shape of the correlation function, it is not possible to fully separate the BAO feature from the rest {of the} information included in the correlation function, i.e. the rescaling by the sound horizon can be due to other reason than the shift of BAO peak. Nevertheless, it is still a good approximation since the BAO is a key feature in the correlation function.

\subsection{Covariance matrices and mock catalogues}
\label{subsec: covmat and EZ}

We estimate the covariance matrices of the measurements described in 
Section~\ref{subsec: wedges and multipoles} using $1\,000$ 
mock catalogues constructed using {\sc EZmocks} \citep{Chuang-2015}.
These simulations are based on initial conditions generated using the Zel’dovich approximation (ZA)\citep{Zeldovich-1970}, with parameters to effectively account for nonlinearities and bias. The probability density function (PDF) of halos is calibrated by mapping the density field to the BigMultiDark (BigMD) $N$-body simulations \citep{Klypin-2016}. Additional scattering is added to the PDF to account for the stochastic bias and a further fitting of the power spectrum and bispectrum is applied to account for nonlinear effects and deterministic bias. 
The bias and Finger of God (FoG) parameters \citep{Kaiser-1987} are calibrated against the DR14 LSS quasar catalogue, with independent treatment for the NGC and SGC. The quasars are assigned directly to the simulated dark matter particles. 
The light-cone mock catalogues are built using seven redshift shells, each of which is taken from a box with size of $(5h^{-1}{\rm Gpc})^3$. All redshift shells for the $i$th mock have the same initial Gaussian density field but with different EZ-parameters. The redshift evolution of the EZ-parameters is determined by solving a system of equation and equating them with the parameters measured from the data within three overlapped redshift bins. The details can be found in \citep{Ata-2017}.
The redshift error is encoded in the EZmcoks intrinsically due to the bias calibration with respect to the real data 

Each EZmocks corresponds to an independent realization of a flat $\Lambda$CDM cosmology 
defined by a matter density parameter $\Omega_{\rm{m}}=0.307$, a baryon density of 
$\Omega_{\rm{b}}h^2=0.022$, a dimensionless Hubble parameter $h=0.678$, and no contribution from massive neutrinos.
The power spectrum of these mocks is characterized by a scalar spectral index $n_{\rm  s}=0.96$, normalized to a value of $\sigma_8(z=0)=0.8225$. These parameters correspond to a value of $f\sigma_8(z=1.52)=0.378$ at the mean redshift of the LSS quasar sample.

\begin{figure}
\centering
\includegraphics[width=7cm]{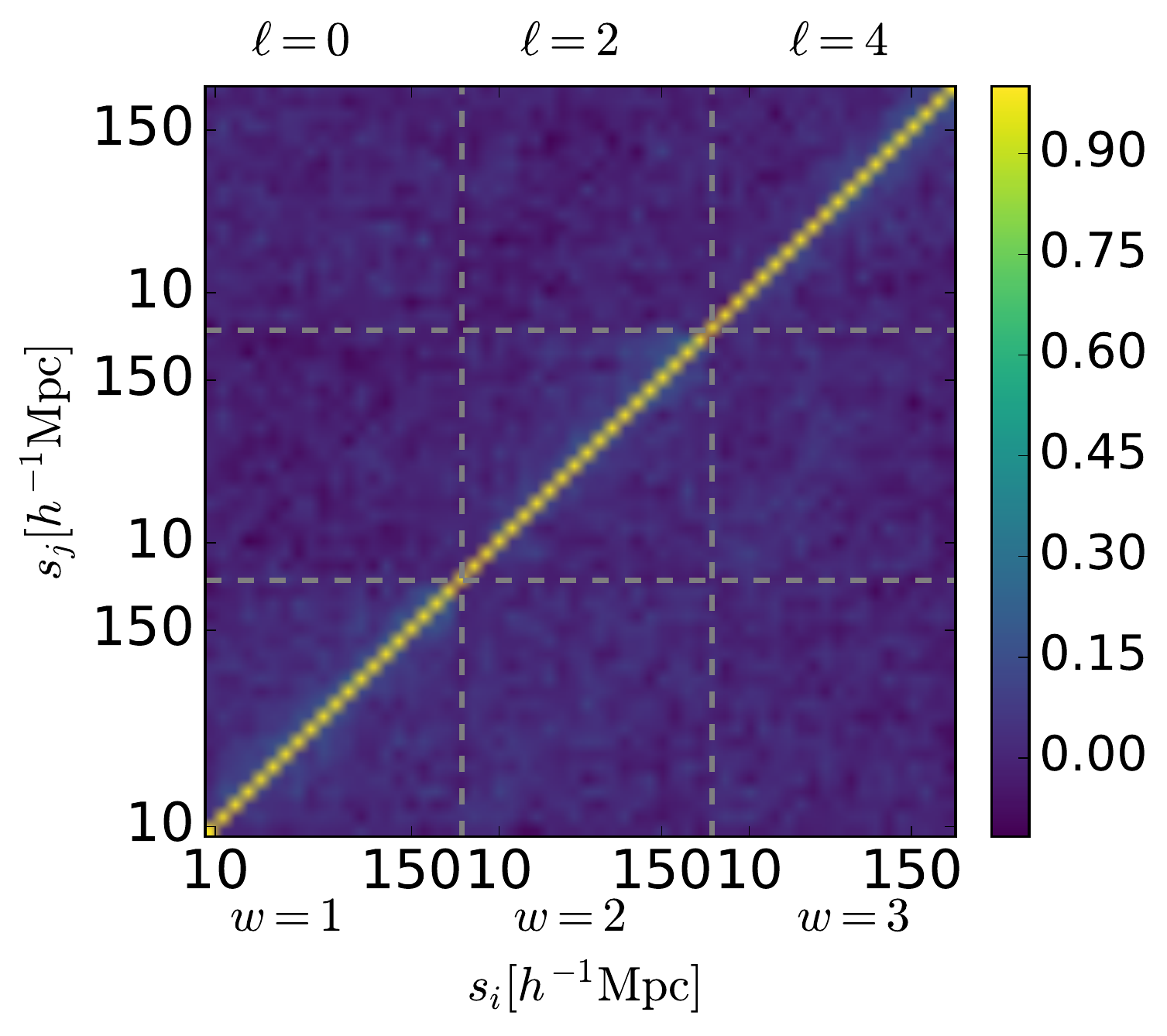}
\caption{Correlation matrices estimated using from our $1\,000$ EZmocks mock catalogues. 
The upper triangle shows the results corresponding to the Legendre multipoles $\xi_{\ell=0,2,4(s)}$ while the lower triangle is the one for three clustering wedges $\xi_{\rm 3w}(s)$.}
\label{fig: corr_EZ}	   
\end{figure}

We computed the Legendre multipoles and wedges of each mock catalogue using the same bin size and weights as for the real eBOSS LSS quasar sample, but assuming the true cosmology of the EZmock runs as our fiducial cosmology. 
These measurements were used to obtain an estimate of the full covariance matrix, $\mathbfss{C}$, 
associated with our clustering measurements, which were rescaled by a factor $1.03$ to account for a mismatch in the number of objects in the mocks and the real eBOSS data. 
Fig.~\ref{fig: corr_EZ} shows the correlation matrices estimated from the EZmocks. The upper 
triangle shows the correlation matrix for the Legendre multipoles and the lower triangle presents the one for three clustering wedges. As expected for covariance matrices with a large shot-noise contribution, the corresponding correlation matrices are dominated by the diagonal elements. 

In addition to the EZmocks, we have also used a small set of high-fidelity mocks constructed from the OuterRim \citep{Habib-2016}, a high-resolution $N$-body simulation characterized by a cubic box of size
$L=3 h^{-1}{\rm Gpc} $ evolving $10240^3$ dark matter particles with a force resolution of $6 h^{-1}{\rm kpc}$ and a mass resolution per particle $m_{\rm p}=1.82 \times 10^9 h^{-1} {\rm M}_{\odot}$. The mocks are built from a single snapshot at $z=1.433$ and based on a (5+1)-parameter Halo Occupancy Distribution model \citep[HOD,][]{Tinker-2012}), where the additional parameter is necessary for modelling the quasar duty cycle. The concentration of each halo is a function of its mass following the prescription detailed in \citep{Ludlow-2014}. The position and velocity of the satellites follow a NFW profile \citep{Navarro-1996}. Three configurations of satellite fraction were constructed for the HOD OuterRim: ${\rm f_{nosat}}=0\%$, ${\rm f_{fsat}}=13\%$ and ${\rm f_{high}}=25\%$, respectively. A Gaussian smearing was applied to each configuration, to mimic the redshift error. The fiducial cosmology for OuterRim is consistent with WMAP7 \citep{Komatsu-2011}, i.e. $\Omega_{\rm m}=0.265$, $\Omega_{\rm b}h^2=0.0235$ $h=0.678$, $\sigma_8=0.8$, $n_s=0.963$ and zero neutrino mass. Further details for OuterRim simulation could be found in \citep{Zarrouk-2017, Gil-Marin-2017}.

\subsection{The likelihood function}
\label{sec: likelihood}

We use Bayesian statistics to infer our cosmological constraints. 
Assuming the evidence of the data is normalized to one, the posterior is given by 
$\mathcal{P}({\bm \lambda}|{\bm \xi})\propto \mathcal{L}({\bm \xi}|{\bm \lambda})\mathcal{P}({\bm \lambda})$, with ${\bm \lambda}$ being the cosmological parameters of interest and a set of nuisance parameters that enter our model (see Section~\ref{subsec: model of CorrFunc}), and ${\bm \xi}$ representing an array containing our clustering measurements.
Assuming Gaussian-distributed data, the likelihood function is,
\begin{equation}
\mathcal{L}({\bm \xi}|{\bm \lambda})\propto \exp\left[-\frac{1}{2}\left({\bm \xi}-{\bm \xi}_{\rm model}({\bm \lambda})\right)^{\rm T}{\rm \Psi}\left({\bm \xi}-{\bm \xi}_{\rm model}({\bm \lambda})\right)\right],
\label{eqn: gaussian-likelihood}
\end{equation}
where ${\rm \Psi}=\mathbfss{C}^{-1}$ and ${\bm \xi}_{\rm model}({\bm \lambda})$ represents 
the theoretical model used to describe our measurements for the parameters included in ${\bm \lambda}$.
As described in Section~\ref{subsec: covmat and EZ}, we estimate the covariance matrices of
our measurements from the sample variance of a set of $1\,000$ mock catalogues. 
The noise in this estimate of $\mathbfss{C}$ makes its inverse a biased 
estimate of ${\bm \Psi}$.  
This can be corrected by including a prefactor in the estimate of the precision matrix 
as \citep*{kaufman-1967,Hartlap-2007}
\begin{equation}
	\hat{\bm{\Psi}}=\left(1-\frac{N_{\rm b}+1}{N_{\rm m}-1}\right)
	\hat{\mathbfss{C}}^{-1},
	\label{eqn: hartlap}
\end{equation}
where $N_{\rm b}$ represents the number of bins in the data vector and $N_{\rm m}$ corresponds to the number of mocks used to estimate $\hat{\mathbfss{C}}$. 
Although unbiased, the estimate of equation~(\ref{eqn: hartlap}) remains affected by noise 
due to the finite number of mock catalogues, which should be propagated into the obtained constraints, increasing the parameter uncertainties \citep{Dodelson-2013, Taylor-2013, Taylor-2014}.
As described in \citet{Percival-2014}, the results obtained when the estimate $\hat{\bm{\Psi}}$ 
is used to compute the Gaussian likelihood function of equation~(\ref{eqn: gaussian-likelihood})
can be corrected to account for this additional uncertainty by rescaling the obtained parameter 
covariances by a factor that depends on $N_{\rm b}$, $N_{\rm m}$, and the dimension of the 
parameter space explored in the analysis, $N_{\rm p}$. However, this simple rescaling does not provide a corrected version of the full parameter posterior distribution $\mathcal{P}({\bm \lambda}|{\bm \xi})$.

\citet{Sellentin-2016} followed a different approach, by marginalising 
equation~(\ref{eqn: gaussian-likelihood}) over the true covariance matrix, conditioned on its estimated value. This procedure leads to a likelihood 
function that deviates from the simple Gaussian recipe, and follows a modified version 
of the multivariate t-distribution given by
\begin{equation}
\mathcal{L}({\bm \xi}|{\bm \lambda})\propto  \left[1+\frac{\left({\bm \xi}-{\bm \xi}_{\rm model}({\bm \lambda})\right)^{\rm T}
\hat{\mathrm{C}}^{-1}\left({\bm \xi}-{\bm \xi}_{\rm model}({\bm \lambda})\right)}{N_{\rm m}-1}\right]^{-\frac{N_{\rm m}}{2}},
\label{eqn: mtdist-likelihood}
\end{equation}
which depends explicitly on the number of mocks on which the estimate $\hat{\mathrm{C}}$ is based. 
The results obtained by sampling this modified likelihood function
correctly account for the additional uncertainty due to the noise in $\hat{\mathrm{C}}$, without the need to include any additional rescaling factor. We use the non-Gaussian likelihood function of equation~(\ref{eqn: mtdist-likelihood}) in our analysis. As discussed in Appendix~\ref{appendix: likelihood}, for the number of mock catalogues used in our analysis, the results obtained by means of this likelihood function and those inferred using the standard Gaussian recipe are essentially identical.

\section{The model}
\label{sec: the_model}

\subsection{Modelling anisotropic clustering measurements}
\label{subsec: model of CorrFunc}

We base the theoretical description of our clustering measurements on a model of the  
power spectrum $P(\mu, k)$, which we Fourier transform to obtain the anisotropic two-point 
correlation function as
\begin{equation}
\xi(\mu,s)=\frac{1}{(2\pi)^3}\int P(\mu,k) e^{i{\bf k}\cdot{\bf s}}\,{\rm d}^3k.
\end{equation}
We adopt the same model of non-linearities, bias, and redshift-space distortions as in 
the analyses of the final BOSS galaxy samples of \citet{Sanchez-2017a}, \citet{Grieb-2017a}, and \cite{Salazar-Albornoz-2017}, which we extend to include the effect of non-negligible 
redshift errors. As this model has been discussed and tested in detail in these analyses, we will only briefly summarize it here.

The starting point of our model is the treatment of the non-linear evolution of the density field. On 
large scales the evolution of density perturbations is determined by cold dark matter; for this we 
use renormalized perturbation (RPT) first proposed in \citet{Crocce-2006} supplemented by imposing 
Galilean invariance (gRPT, Crocce, Blas and Scoccimarro in prep.). To describe the clustering of the quasar 
sample we follow \citet{Chan-2012b} and parametrize the bias relation between the matter density fluctuations 
$\delta$ and the quasar density fluctuations, $\delta_{\rm g}$, as
\begin{equation}
\begin{split}
\delta_{\rm g} & =  b_1 \delta + {b_2 \over 2} \delta^2 + \gamma_2\, {\cal G}_2  + \gamma_3^- \, \Delta_3{\cal G}
+ \ldots
\end{split}
\label{eqn:bias}
\end{equation}
where $b_1$ and $b_2$ are the standard linear and quadratic bias \citep{Fry-1993} and the only cubic term 
that enters into the one-loop propagator in the RPT description of bias 
\citep[same as for nonlinear evolution,][]{Crocce-2006, Bernardeau-2008, Bernardeau-2012} has been written 
down. The non-local bias terms $\gamma_{\,2}$ and $\gamma_{\,3}^{-}$ represent the amplitude of the 
Galileon operators of normalized density and velocity potentials, $\Phi$ and $\Phi_{v}$,
\begin{align}
&{\cal G}_2(\Phi_v) = (\nabla_{ij}\Phi_v)^2-(\nabla^2\Phi_v)^2, \\
&\Delta_3{\cal G} =  {\cal G}_2(\Phi)- {\cal G}_2(\Phi_v).
\label{eqn: Galileons}
\end{align}
Under the assumption of local-Lagrangian bias, the non-local bias parameters are determined by 
the linear bias $b_1$ as
\begin{align}
\gamma_2&=-\frac{2}{7}(b_1-1), \label{eqn: nonlocal-bias_2}\\
\gamma_3^{-}&=-\frac{11}{42}(b_1-1).
\label{eqn: nonlocal-bias_3}
\end{align}

Using these ingredients, we describe the redshift-space power spectrum as
\begin{equation}
P(k,\mu)=F_{\rm FOG}(k,\mu)\,P_{\rm novir}(k,\mu)\,{\rm exp}\left[-\left(k\mu\sigma_{\rm zerr}\right)^2\right].
\label{eqn: tot-model}
\end{equation}
$P_{\rm novir}(k,\mu)$ represents the ``no-virial'' power spectrum, given by the 
sum of three contributions
\begin{align}
P_{\rm novir}(\mu,k)=&P_{\rm novir}^{(1)}(k,\mu)+(k\mu f)P_{\rm novir}^{(2)}(k,\mu)\\
&+(k\mu f)^2P_{\rm novir}^{(3)}(k,\mu),
\label{eqn: PS-nonvir}
\end{align}
where
\begin{align}
P_{\rm novir}^{(1)}(k,\mu)&=P_{\rm gg}+2f\mu^2P_{{\rm g}\theta}+f^2\mu^4P_{\theta\theta},\label{eqn: nlkaiser}\\
P_{\rm novir}^{(2)}(k,\mu)&=\int \frac{d^3p}{(2\pi)^3} \frac{p_z}{p^2}\left[B_{\sigma}({\bf p,k-p,-k})-B_{\sigma}(\bf{p,k,-k-p})\right],
\label{eqn: A-trispectrum}\\
P_{\rm novir}^{(3)}(k,\mu)&=\int \frac{d^3p}{(2\pi)^3} F({\bf p})F(\bf{k-p}).
\label{eqn: B-quad-linear}
\end{align}
Here, $P_{\rm novir}^{(1)}(k,\mu)$ corresponds to a non-linear version of the Kaiser formula \citep{Kaiser-1987}, and $P_{\rm novir}^{(2)}(k,\mu)$ and $P_{\rm novir}^{(3)}(k,\mu)$ are given by tree-level bispectrum and quadratic linear-theory power spectrum.

The modelling of the RSD is based on \citep{Scoccimarro-2004}. Eqn. 
(\ref{eqn: PS-nonvir}) includes the distortion of BAO on large scales, while on small scales the random motion of LSS smears the distribution along the line of sight direction and give rise to the FoG effect, 
\begin{equation}
F_{\rm FOG}(\mu,k)\equiv \frac{1}{\sqrt{1+f^2\mu^2k^2a_{\rm vir}^2}}\exp\left(\frac{-f^2\mu^2k^2\sigma_v^2}{1+f^2\mu^2k^2a_{\rm vir}^2}\right),
\end{equation}
with $a_{\rm vir}$ being a free parameter that represents the kurtosis of the small scale 
velocity distribution. For the analysis in this paper the velocity dispersion $\sigma_v$ is 
calculated from a linear theory prediction and is treated as scale-invariant.

{
Given the high redshift quasar sample, the uncertainties in redshift estimates are larger compared to the galaxies and they can as well be redshift dependent \citep{Dawson-2016}. The uncertainty in the redshift estimates can have impact on the small scale clustering. We use a simple model by approximating it as a Gaussian damping to the power spectrum \citep{Blake-2005}. A global $\sigma_{\rm zerr}=c\delta z/H(z_{\rm eff})$ is fitted at the effective redshift shift $z_{\rm eff}=1.52$, where $c$ is the speed of light,
}
and $\delta z$ is the uncertainty in determining the redshift. The uncertainty in determining the radial distance of a given object can be translated into the velocity dispersion in unit of ${\rm km s^{-1}}$,
\begin{equation}
\Delta v= \frac{\delta z}{1+z}c = \frac{\sigma_{\rm zerr} H(z_{\rm eff})}{1+z} .
\label{eqn: vel_smearing}
\end{equation}
{
This simplified treatment of the redshift error does not reproduce the true evolution of the redshift uncertainty of the eBOSS quasar sample
shown in Fig. 7 of \citet{Dawson-2016}.
However, as we will see in sec. \ref{subsec: model test}, this ansatz allows us to recover unbiased cosmological parameters after marginalizing over the $\sigma_{\rm zerr}$.}
In summary, our full model of $P(k,\mu)$ is characterized by six nuisance parameters, the bias factors $b_1$, $b_2$, $\gamma_2$, and $\gamma_3^-$, the FoG parameter $a_{\rm vir}$, and the redshift error $\sigma_{\rm zerr}$. However, as described in the next section, the eBOSS quasar sample cannot constrain the non-local bias parameters $\gamma_2$ and $\gamma_3^-$, which we then set according to the local-Lagrangian relations of equations~(\ref{eqn: nonlocal-bias_2}) and (\ref{eqn: nonlocal-bias_3}). The remaining parameters are treated as free quantities and marginalized over in our analysis. 

\subsection{Model validation}
\label{subsec: model test}

The model described in sec. \ref{subsec: model of CorrFunc} was tested in detail for the analyses of the final BOSS galaxy samples \citep[see][]{Sanchez-2017a,Grieb-2017a,Salazar-Albornoz-2017}.
We focus here on testing the modelling of the impact of non-negligible redshift errors. 
We employ our tests on the same set of EZmocks synthetic catalogues described in 
Section~\ref{subsec: covmat and EZ}, on which we base our estimates of the covariance matrices
of our measurements. 

\begin{figure*}
\centering
\includegraphics[width=\columnwidth]{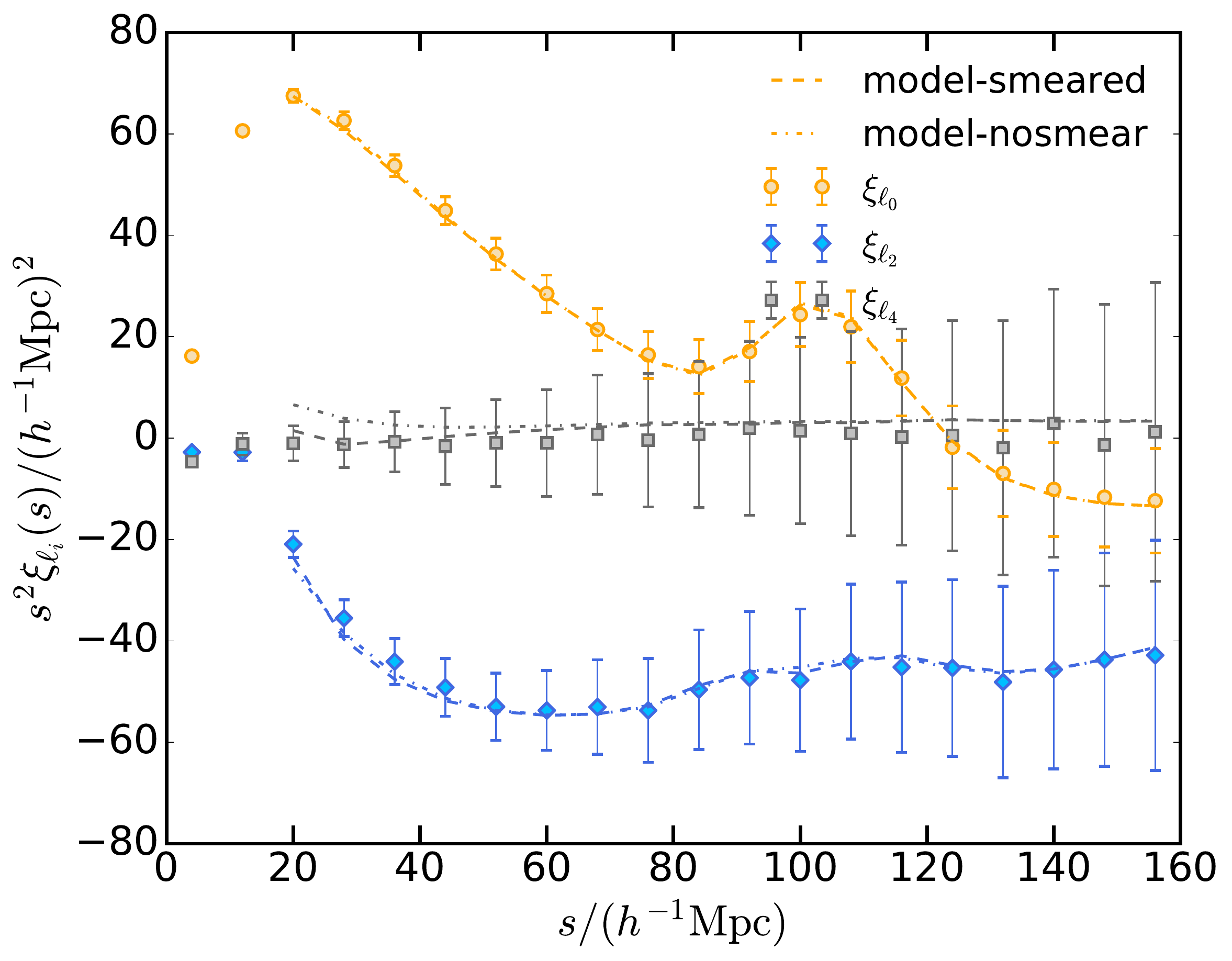}
\includegraphics[width=\columnwidth]{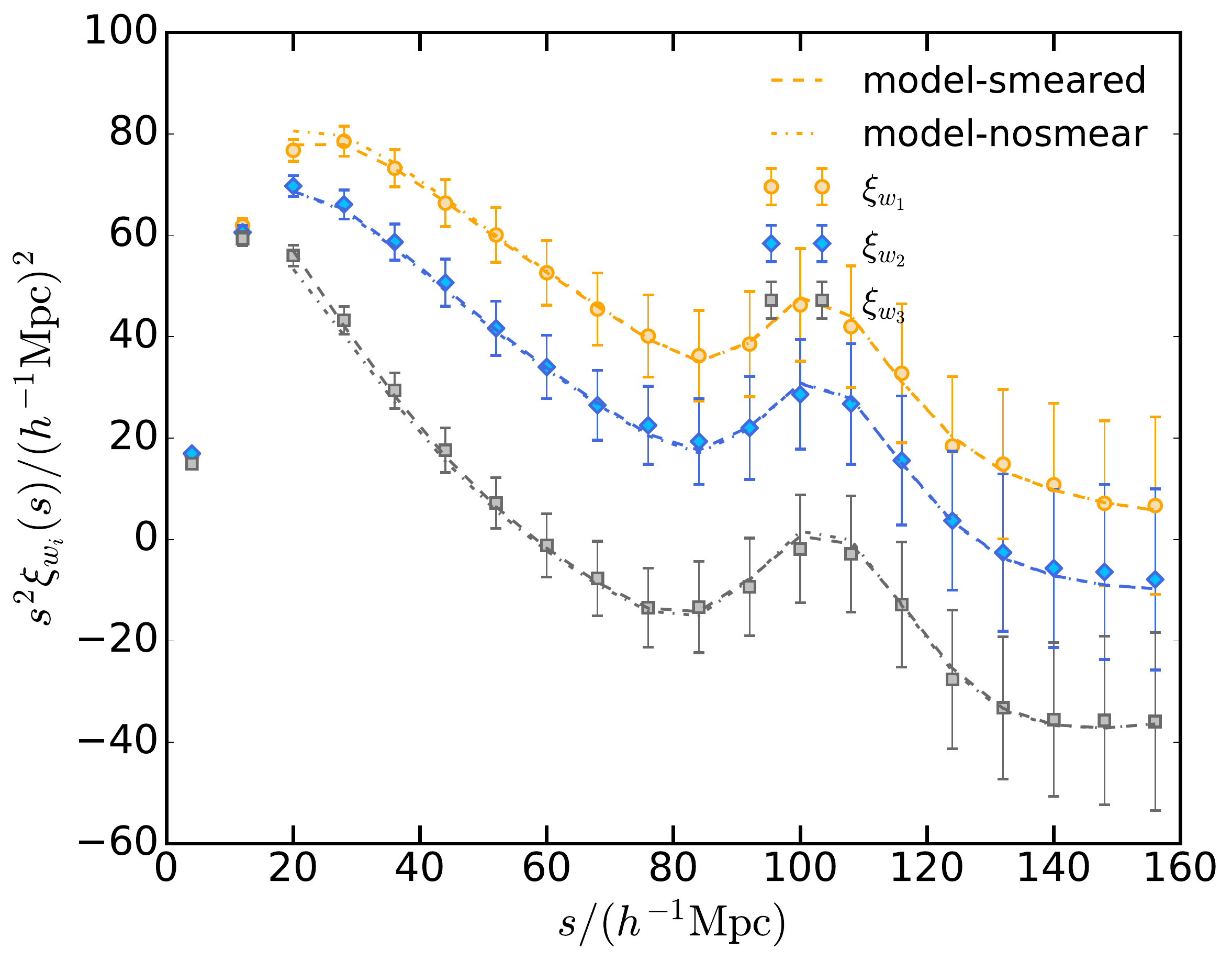}
\caption{Left: Legendre multipoles, monopole(red), quadrupole (cyan), and hexadecapole (grey). Right: clustering wedges in the directions parallel (red) intermediate (cyan) and transverse (grey) to the line of sight measured from the EZmocks. The dashed lines correspond to the best fitting model to these measurements including the redshift error parameter, labeled as "smeared". The dash-dotted lines corresponds to the same model but without the redshift error parameter, labelled as "nosmear". The errorbars are inferred from $10^3$ sets of mock catalogues (EZmocks).}
\label{fig: model_test_EZ}	   
\end{figure*}

The points in Fig.~\ref{fig: model_test_EZ} correspond to the mean Legendre Multipoles 
(left panel) and clustering wedges (right panel) of the EZmocks. The error bars are obtained 
from the square root of the diagonal terms of the covariance matrix estimated from the same set of mocks. We tested our model by performing fits to these measurements using our model for various configurations in order to assess its ability to recover unbiased constraints.

As a first test, we fixed the values of all cosmological parameters to the correct values for the cosmology of the mocks and varied only the nuisance parameters $b_1$, $b_2$, $a_{\rm vir}$, and $\sigma_{\rm zerr}$. Given the volume and number density of the quasar LSS sample, and hence of the EZmocks, the values of the non-local bias parameters cannot be constrained by the data. 
{
We performed tests with or without varying the non-local bias parameters, and found that it has no impact on the obtained constraints or the quality of the fits.
}
We therefore opted for setting their values in terms of $b_1$ according to the local-Lagrangian predictions of equations~(\ref{eqn: nonlocal-bias_2}) and (\ref{eqn: nonlocal-bias_3}).  
The dashed lines in Fig.~\ref{fig: model_test_EZ} correspond to the best-fitting
models obtained when the redshift error $\sigma_{\rm zerr}$ is treated as a free parameter and included in the fits, while the dot-dashed lines represent the results obtained when setting $\sigma_{\rm zerr}=0$. Although both models provide a good description of the mock measurements, the results obtained when $\sigma_{\rm zerr}$ is allowed to vary provide a slightly better fit on scales $20\ h^{-1} {\rm Mpc}\leq s\leq 40\ h^{-1} {\rm Mpc}$, as well as at the BAO feature.

As a further test of our model, the parameters $q_{\perp}$, $q_{\parallel}$, and $f\sigma_8$ were 
allowed to vary alongside the nuisance parameters of the model. 
Table~\ref{tab: param_space_nuisance} presents a summary of the full set of parameters 
explored in this case. We used flat priors for all parameters, with a 
uniform distribution within the limits specified in the same table. 
The parameter space was explored by means of the Monte Carlo Markov Chain (MCMC) 
technique, applying Metropolis-Hastings algorithm \citep{Metropolis-1953, Hastings-1970}. 
We adopted a Gelman-Rubin criteria \citep{Gelman-1992} of $\hat{R}-1<0.02$ as a measure
of the convergence of the chains.

\begin{figure}
\centering
\includegraphics[width=8cm]{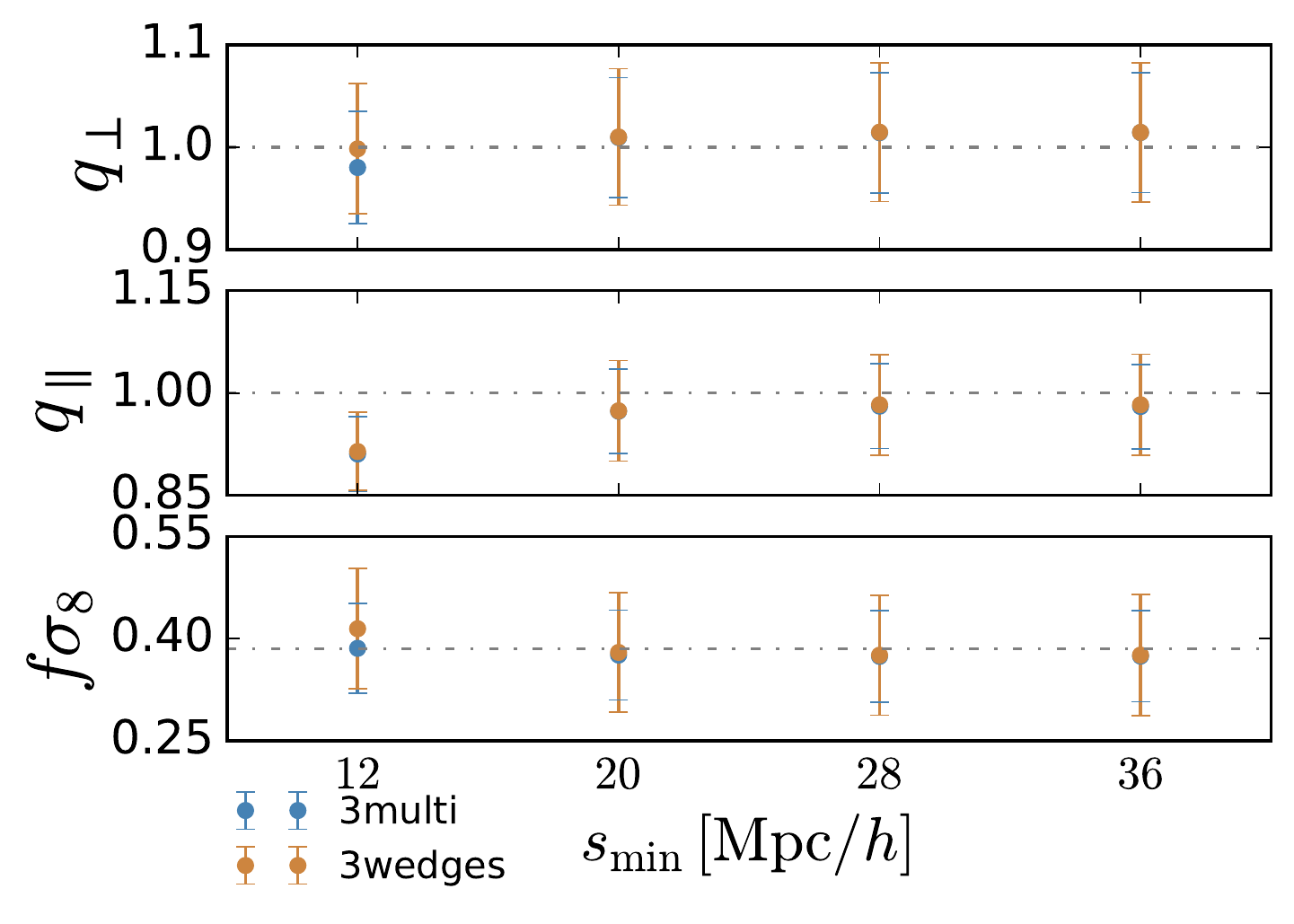}
\caption{Test of the stability of the parameter constraints by varying the fitting range 
$\Delta s=s_{\rm min}-156 \, [{\rm Mpc}/h]$ 
of the mean Legendre multipoles $\xi_\ell(s)$ with $\ell=0,2,4$ (blue) and 
three clustering wedges $\xi_{3w}(s)$ (orange) measured from EZmocks.  
The different panels indicate the marginalized constraints on $q_{\perp}$, $q_{\perp}$ and 
$f\sigma_8$. The dashed lines show the true parameter values for the cosmology of the EZmocks.}
\label{fig: EZ_mean_rangefit}	   
\end{figure}

Fig. \ref{fig: EZ_mean_rangefit} shows the stability of the constraints obtained as a function of the minimum scale included in the fits, for 
$12\, h^{-1}{\rm Mpc}\leq s_{\rm min} \leq 36 \,h^{-1}{\rm Mpc}$. 
These tests were performed 
using the mean of the Legendre multipoles $\xi_{\ell}(s)$ with $\ell=0,2,4$ (blue dots) 
and three clustering wedges $\xi_{\rm 3w}(s)$ (orange dots) measured from the mocks. 
The dashed lines in each panel correspond to the true parameter values for the cosmology of the mocks. The constraints on $q_{\perp}$, $q_{\parallel}$, and $f\sigma_8$ are stable over 
different minimum fitting range, with multipoles and wedges providing results in good 
agreement over the full range of values of $s_{\rm min}$ considered in this test. 
However, the results obtained from $\xi_\ell(s)$ possess slightly smaller uncertainties.
Based on these tests, we defined a minimum scale of $s_{\rm min}=20\, h^{-1}{\rm Mpc}$ 
for our fits to the true eBOSS data. 
{
\citet{Chuang-2015b} tested the accuracy of EZmocks against the BigMD full N-body simulation. This comparison shows that the accuracy of the monopole measured from the EZmocks varies from $1\%-5\%$ down to $10 \, h^{-1}{\rm Mpc}$ scales, depending on the halo finder. The quadrupole reaches $10\%-15\%$ precision for scales $s\sim 10 \, h^{-1}{\rm Mpc}$. Therefore, these mocks give an accurate description of the clustering properties on the scales used in our analysis.}

\begin{table}
\centering
\caption{A summary of the parameter space $\lambda$. A flat prior is applied to all parameters with uniform distribution inside the limits and zero otherwise.}	
\begin{tabular}{ | l | l | l | c |  }
\hline
Parameter & Description & Units &  Prior limits   \\ 
\hline
$b_1$ & Linear bias &  $-$ & $[0.25, 6]$ \\
$b_2$ & Second order bias & $-$ &[$-1, 6]$ \\
$a_{\rm vir}$ & FoG kurtosis & $-$ & $[0.2, 5]$ \\
$\sigma_{\rm zerr}$ & Redshift error & ${\rm Mpc}/h$ & $[0, 6]$\\
\hline
$q_{\perp}$ &  Distortion ${\perp}$ L.O.S&  $-$ & $[0.5, 1.5]$ \\
$q_{\parallel}$ & Distortion ${\parallel}$ L.O.S &  $-$ & $[0.5, 1.5]$ \\
$f\sigma_8$ & growth parameter &  $-$ & $[0, 1]$ \\
\hline
\end{tabular}
\label{tab: param_space_nuisance}
\end{table}

\begin{figure*}
\centering
\includegraphics[width=14cm]{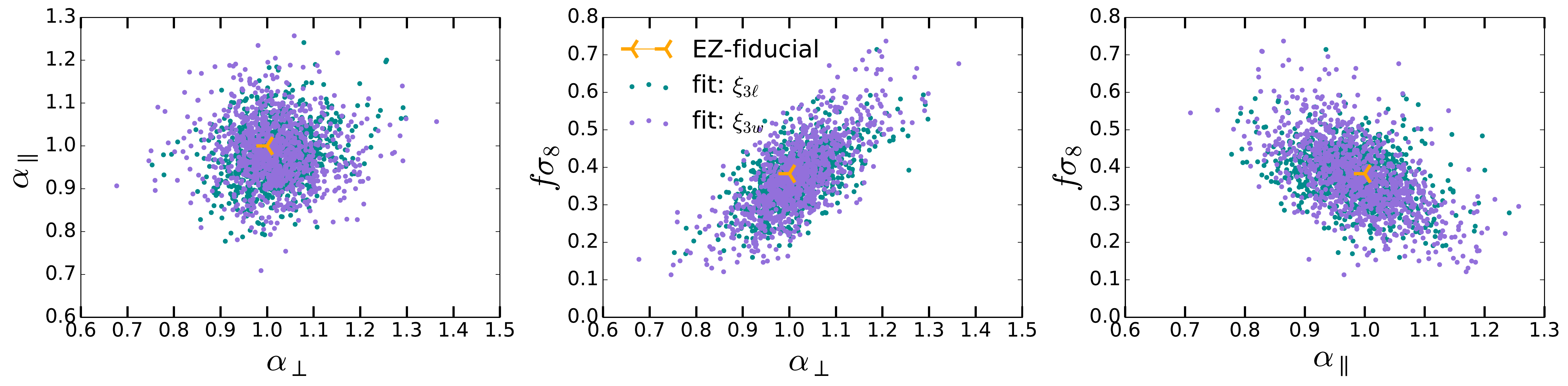}
\includegraphics[width=14cm]{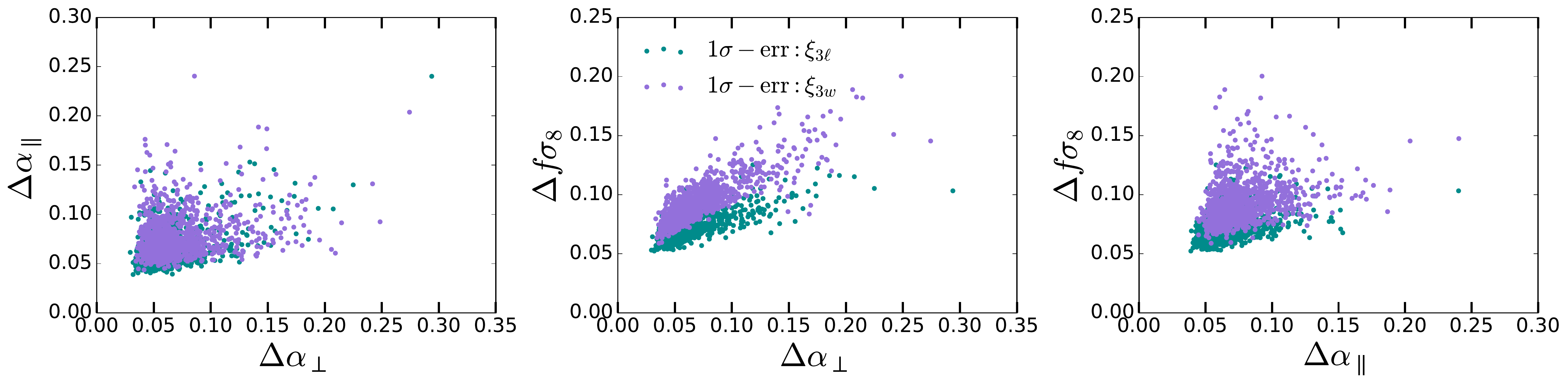}
\caption{Upper panels: constraints on $\alpha_{\perp}$, $\alpha_{\parallel}$, and 
$f\sigma_8$ obtained when fitting the Legendre multipoles $\xi_{\ell=0,2,4}(s)$ 
(cyan) and three clustering wedges $\xi_{\rm 3w}(s)$ (purple) of each mock 
catalogue. 
The orange cross in the center of each panel represents the values corresponding to the 
true cosmology of the EZmocks. 
Lower panel: 68\% confidence levels on $\alpha_{\perp}$, $\alpha_{\parallel}$, and 
$f\sigma_8$ inferred from the fits to the Legendre multipoles (grey) and clustering wedges 
(brown). }
\label{fig: 2d_param_EZ}	   
\end{figure*}

Using this range of scales, we performed fits to the measurements of $\xi_{\ell=0,2,4}(s)$ 
and $\xi_{3w}(s)$ obtained from each mock catalogue. 
The upper panels of Fig. \ref{fig: 2d_param_EZ} present the mean values of $\alpha_{\perp}$, 
$\alpha_{\parallel}$, and $f\sigma_8$ obtained from the fits to Legendre multipoles (cyan points) 
and clustering wedges (purple) of the individual EZmocks. 
The lower panels of the same figure show the symmetrised 68\% uncertainties on these 
parameters recovered from the MCMC fits for clustering wedges (brown  points) and Legendre 
multipoles (grey points). 
This comparison also demonstrates that the Legendre multipoles $\xi_{\ell=0,2,4}(s)$ provide on average slightly tighter constraints than the measurements of $\xi_{\rm 3w}(s)$. 
As we will see in Section~\ref{sec: constraints} this behaviour is also the case for our fits to the real eBOSS quasar clustering measurements. 

\begin{figure}
\centering
\includegraphics[width=8.5cm]{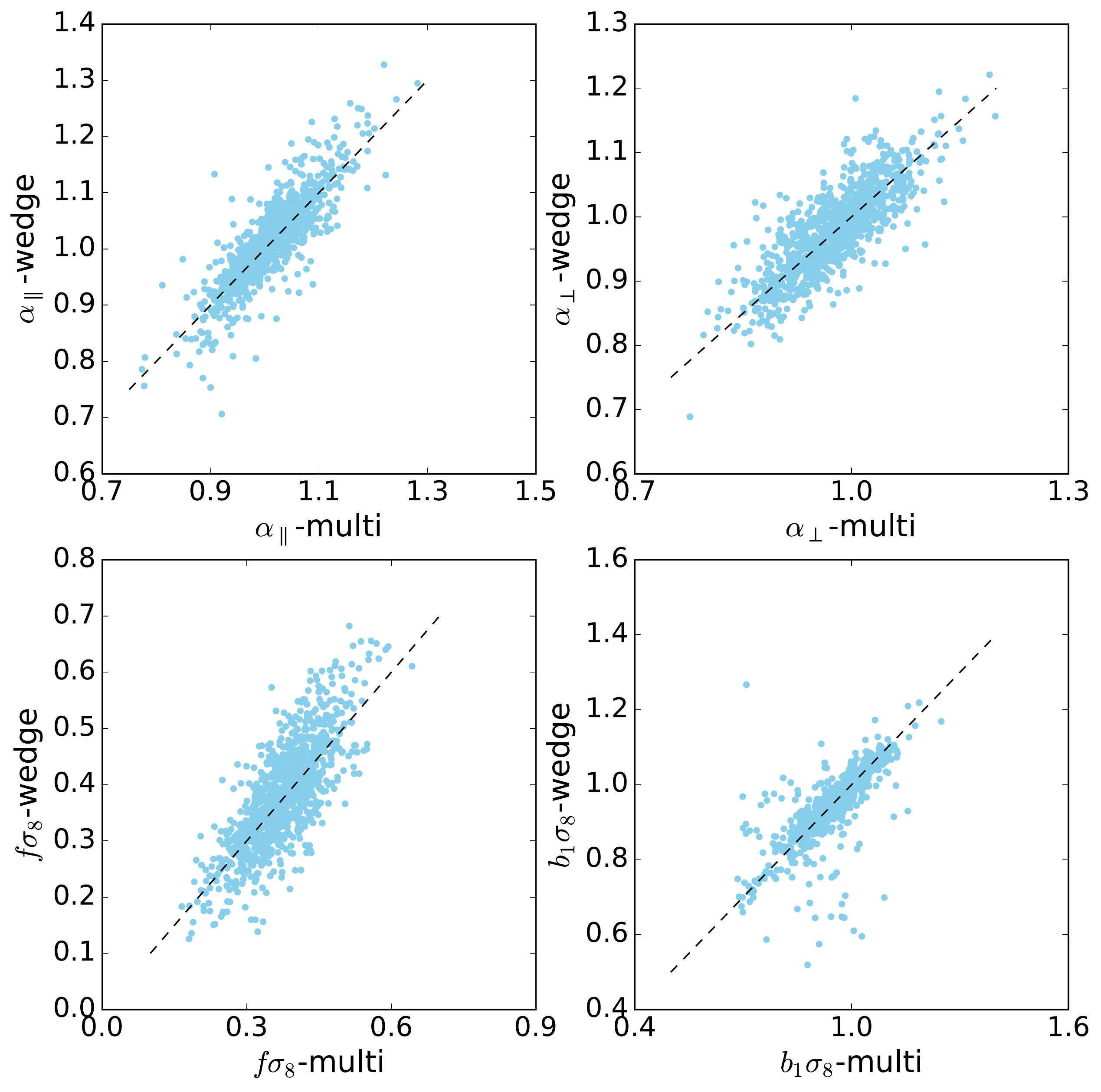}
\caption{Comparison of the constraints on $\alpha_{\perp}$, $\alpha_{\parallel}$, 
$f\sigma_8$ and $b_1\sigma_8$ obtained from the analysis of Legendre multipoles ($x$ axis)
and three clustering wedges ($y$ axis) of each of our mock catalogues. The dashed line 
corresponds to a one-to-one relation.}
\label{fig: 2d_corr_EZ}	   
\end{figure}

Fig. \ref{fig: 2d_corr_EZ} shows the mean values of $\alpha_{\perp}$, 
$\alpha_{\parallel}$, $f\sigma_8$, and $b\sigma_8$ inferred from each individual mock using 
multipoles and wedges. The results obtained from these statistics are completely consistent with each other, with correlation coefficients close to one. 
The scattering in the panels is due to the fact that multipoles and wedges pick slightly different information from the two-dimensional correlation function. 
The relation between the values of $f\sigma_8$ deserves special attention. These results show a lower correlation between the fits to $\xi_{\ell=0,2,4}(s)$ and $\xi_{\rm 3w}(s)$ than in the other cases. This behaviour is due to the larger scatter in the values of $f\sigma_8$ obtained from the clustering wedges. 

\begin{table}
\centering
\caption{Parameter constraints for $\alpha_{\perp}$, $\alpha_{\parallel}$ and $f\sigma_8$ 
derived from the fit to individual $10^3$ of EZmocks using clustering wedges and Legendre multipoles. The errors are derived from the scattering of the mean value for fitting each of the chains, with fiducial value $f\sigma_8(z=1.52)=0.378$. The fitting range is ${\rm ds}=20 h^{-1}{\rm Mpc} -156 h^{-1}{\rm Mpc}$. The effect of fixing the redshift error $\sigma_{\rm zerr}=0$ can be seen on the second column.}	
\begin{tabular}{ | l | c | c | c |  }
\hline
Statistic & Parameter & $\sigma_{\rm zerr}\neq 0$ &  $\sigma_{\rm zerr} = 0$  \\
\hline
$\xi_{\ell}(s)$ & $\Delta\alpha_{\perp}$    &$\, 0.010\pm 0.064$  &$\,0.042\pm 0.069$ \\
               & $\Delta\alpha_{\parallel}$ &$-0.026\pm 0.060$  &$-0.068\pm 0.059$ \\
               & $\Delta f\sigma_8$         &$-0.003\pm 0.070$  &$\,0.012\pm 0.076$  \\
               & $\sigma_{\rm zerr}$       &$2.882\pm 0.067$ & $-$\\
\hline
$\xi_{\rm 3w}(s)$    & $\Delta\alpha_{\perp}$     &$\,0.012\pm 0.075$  &$\,0.065\pm 0.083$ \\
               & $\Delta\alpha_{\parallel}$ &$-0.024\pm 0.066$  &$-0.084\pm 0.065$ \\
               & $\Delta f\sigma_8$        &$\,0.003\pm 0.093$  &$\,0.057\pm 0.113$  \\
               & $\sigma_{\rm zerr}$       &$2.873\pm 0.067$ & $-$\\
\hline
\end{tabular}
\label{tab: sigzerr_test_EZ}
\end{table}

As an illustration of the impact of introducing a non-zero redshift error in our model, we performed 
additional fits to each mock catalogue setting $\sigma_{\rm zerr}=0$.
Table \ref{tab: sigzerr_test_EZ} presents the average difference between the
values recovered from the fits of Legendre multipoles and wedges of each 
EZmocks and their corresponding true values.  
The first set of values corresponds to those recovered when $\sigma_{\rm zerr}$ is varied with the flat 
prior given in table \ref{tab: param_space_nuisance}, while in the second column shows the results assuming 
$\sigma_{\rm zerr} = 0$. We have also tested using larger prior $[0,20]$ on $\sigma_{\rm zerr}$ and the 
resulting changes in the inferred parameters are less than few percent of $\sigma$.
The listed error is inferred from the scatter of the fitted mean value for each individual mock, and hence 
indicates the statistical error that can be expected for the measurements of these parameters based on one 
realization of the eBOSS DR14 quasar LSS sample. 
The comparison of these values show that ignoring the non-negligible redshift errors affecting the 
measurements can potentially bias the obtained constraints, leading to an 
overestimation of $f\sigma_8$ and an underestimation of $\alpha_{\parallel}$ for both 
multipoles and wedges. 
The inferred $\sigma_{\rm zerr}(z=1.52) $ corresponds to a dispersion 
$\sim 180 s^{-1}{\rm km}$ using Eqn. (\ref{eqn: vel_smearing}).
Although the impact of a non-zero redshift error in our model seems marginal from 
Fig. \ref{fig: model_test_EZ}, the deviations between the true and inferred parameter values 
are significantly 
reduced in the case in which $\sigma_{\rm zerr}=0$ is treated as a free parameter, leading 
to systematic 
differences that are much smaller than the expected statistical uncertainties of the eBOSS sample. 

{
Given the wide redshift range of the DR14 quasar sample, representing our results in terms of 
cosmological constraints at an effective redshift needs to be validated.
The possible impact of light-cone effects can be assessed by means of the EZmocks mock 
catalogues, 
which cover the same redshift range as the eBOSS QSO catalogue and take into account 
the redshift evolution of cosmic structure.
The good match between the inferred AP parameters and $f\sigma_8$ with the 
fiducial values of the EZmocks justifies this approximation.}

As a further tests of our model we applied it to the analysis of the 
OuterRim HOD mocks described in Section~\ref{subsec: covmat and EZ}. 
We focus on the analysis of the samples including redshift errors (the {\it smeared} samples), as 
these are the ones that should more closely resemble the characteristics of the real eBOSS quasar 
catalogue. 
We restrict the analysis to the range $0.8 < z < 2$, leading to a mean redshift of 
$z = 1.433$.
As 100 realizations are not enough to compute robust covariance matrices, 
we based our fits on theoretical covariance matrices computed following the Gaussian recipe of \citet{Grieb-2016}, for the volume and mean number density of each HOD sample. Although these simple predictions do not take into account the redshift evolution of the number density of the samples, we have found that a simple rescaling of the theoretical covariances by a factor 1.4 gives a good match to the variance inferred from the 100 realizations. 
Table \ref{tab: sys_test_OR} summarizes the results obtained when fitting the mean of the 
Legendre Multipoles and clustering wedges of the OuterRim HOD mock catalogues. 
We list the difference between the mean parameter values inferred from our fits and their true values. In all cases, the error quoted corresponds to the statistical uncertainty expected for one realization, which are similar to the ones expected for the eBOSS quasar samples. 
Following \citet{Zarrouk-2017}, we limit the maximum scales included in the analysis to 
$s_{\rm max} = 135\,h^{-1}{\rm Mpc}$, but otherwise apply the same set-up as 
in the analysis of the EZmocks. 
In all cases, the recovered values of $f\sigma_8$ are lower than the 
true one for the OuterRim fiducial cosmology, $f\sigma_8(z=1.433)=0.382$.
The cause of this systematic shift is not identified. Further tests of the accuracy of our model of non-linearities, bias and redshift-space distortions at high redshift are required. The details of the implementation o the HOD (placing the central galaxy at the centre of mass of the halo and assuming an NFW distribution for the positions and velocities of the satellites) might also 
play a role in the results \citep[see e.g.][]{Orsi-2017}. 
As described in Section \ref{subsec: BAO-RSD}, we use 
the results inferred from the OuterRim HOD mocks to define a systematic error budget associated with 
our measurements and leave a more detailed analysis of the origin of these differences for future studies.

Based on the tests presented in this section, we adopted the results derived from the analysis of three Legendre multipoles, when the redshift-error parameter $\sigma_{\rm zerr}$ is varied and marginalized over, as our main parameter constraints.

\begin{table}
\caption{Parameter constraints for $\alpha_{\perp}$, $\alpha_{\parallel}$ and $f\sigma_8$ 
derived from the mean of OuterRim using clustering wedges and Legendre multipoles for different satellite fractions ${\rm f_{sat}}$. The errors are derived from the symmetrised $68\%$ percentile with fiducial value $f\sigma_8(z=1.433)=0.382$. The fitting range is ${\rm ds}=20 h^{-1}{\rm Mpc} -135 h^{-1}{\rm Mpc}$. }
\begin{tabular}{ | l | c | c | c | c |  }
\hline
Stat. & Param. & ${\rm f_{\rm sat}}=0\%$ &  ${\rm f_{\rm sat}}=13\%$ & ${\rm f_{\rm sat}}=25\%$  \\
\hline
$\xi_{\rm \ell}(s)$   & $\Delta\alpha_{\perp}$   & $0.018\pm 0.049$  &$-0.002 \pm 0.046$ &$-0.004 \pm 0.040$\\
               & $\Delta\alpha_{\parallel}$ & $0.036\pm 0.066$  &$0.025 \pm 0.063$ &$0.018 \pm 0.054$\\
    & $\Delta f\sigma_8$     & $-0.043\pm 0.072$  &$-0.044 \pm 0.066$  &$-0.030 \pm 0.062$\\
\hline
$\xi_{\rm 3w}(s)$   & $\Delta\alpha_{\perp}$   & $0.015\pm 0.055$  &$-0.001 \pm 0.048$ &$-0.007 \pm 0.043$\\
               & $\Delta\alpha_{\parallel}$ & $0.034\pm 0.076$  &$0.024 \pm 0.068$ &$0.021 \pm 0.057$\\
    & $\Delta f\sigma_8$     & $-0.046\pm 0.080$  &$-0.043 \pm 0.073$  &$-0.033 \pm 0.067$\\

\hline
\end{tabular}
\label{tab: sys_test_OR}
\end{table}

\section{Cosmological implications}
\label{sec: constraints}
In this section we explore the cosmological implications of our clustering measurements. 
In section~\ref{subsec: BAO-RSD} we present
the results obtained by fitting the model of non-linear clustering in redshift space 
described in Section~\ref{subsec: model of CorrFunc} to the measurements of the Legendre 
multipoles and $\mu$-wedges of the eBOSS quasar sample. 
Section~\ref{subsec: companion} compares our results with those of the eBOSS companion papers.

\subsection{BAO and RSD constraints}
\label{subsec: BAO-RSD}

We used the model of two-point clustering described in Section~\ref{subsec: model of CorrFunc} 
to extract the cosmological information contained in the Legendre Multipoles and clustering
wedges of the eBOSS DR14 LSS quasar sample. We followed the same methodology as in the 
tests of Section~\ref{subsec: model test}, i.e., we included scales in the range 
$20 \,h^{-1}{\rm Mpc} \leq s \leq 156 \,h^{-1}{\rm Mpc}$ and fitted for the parameters
$\alpha_{\perp}$, $\alpha_{\parallel}$, and $f\sigma_8(z)$. The nuisance parameters of our model,
$b_1$, $b_2$, $a_{\rm vir}$, $\sigma_{\rm zerr}$, are included in our MCMC and marginalized 
over in our results, while the values of the non-local bias parameters $\gamma_2$ and 
$\gamma^{-}_3$ are set using equations~(\ref{eqn: nonlocal-bias_2}) and 
(\ref{eqn: nonlocal-bias_3}).
We performed analyses of the multipoles $\xi_{\ell=0,2,4}(s)$ and three
clustering wedges $\xi_{\rm 3w}(s)$. For completeness, we also applied our model to 
the monopole-quadrupole pair, and to two wide $\mu$-wedges $\xi_{\rm 2w}(s)$.
The lines in Fig.~\ref{fig: obs_2ptcf_ds8} correspond to the best-fit models.

The constraints on $\alpha_{\perp}$ and $\alpha_{\parallel}$ obtained from these fits 
can be transformed into measurements of the combinations $D_{\rm M}(z)/r_{\rm d}$ and 
$H(z)r_{\rm d}$. Alternatively, these results can be expressed in terms of 
$D_{\rm V}(z)/r_{\rm d}$, where 
\begin{equation}
D_{\rm V}(z)=\left(D_{\rm M}(z)^2\frac{cz}{H(z)}\right)^{1/3},
\end{equation} 
and the Alcock-Paczynski parameter
\begin{equation}
F_{\rm AP}(z) = D_{\rm M}(z)H(z)/c.
\end{equation}
We chose this basis to represent our results which, taking into account also the 
the growth rate, correspond to measurements of the array
\[
{\bf D}=\left(
\begin{array}{c}
    D_{\rm V}(z_{\rm eff})/r_{\rm d} \\
    F_{\rm AP}(z_{\rm eff})  \\
    f\sigma_8(z_{\rm eff})  
\end{array}\right)
\]
at the effective redshift of the quasar LSS sample, $z_{\rm eff}=1.52$.

\begin{figure*}
\centering
\includegraphics[width=5cm]{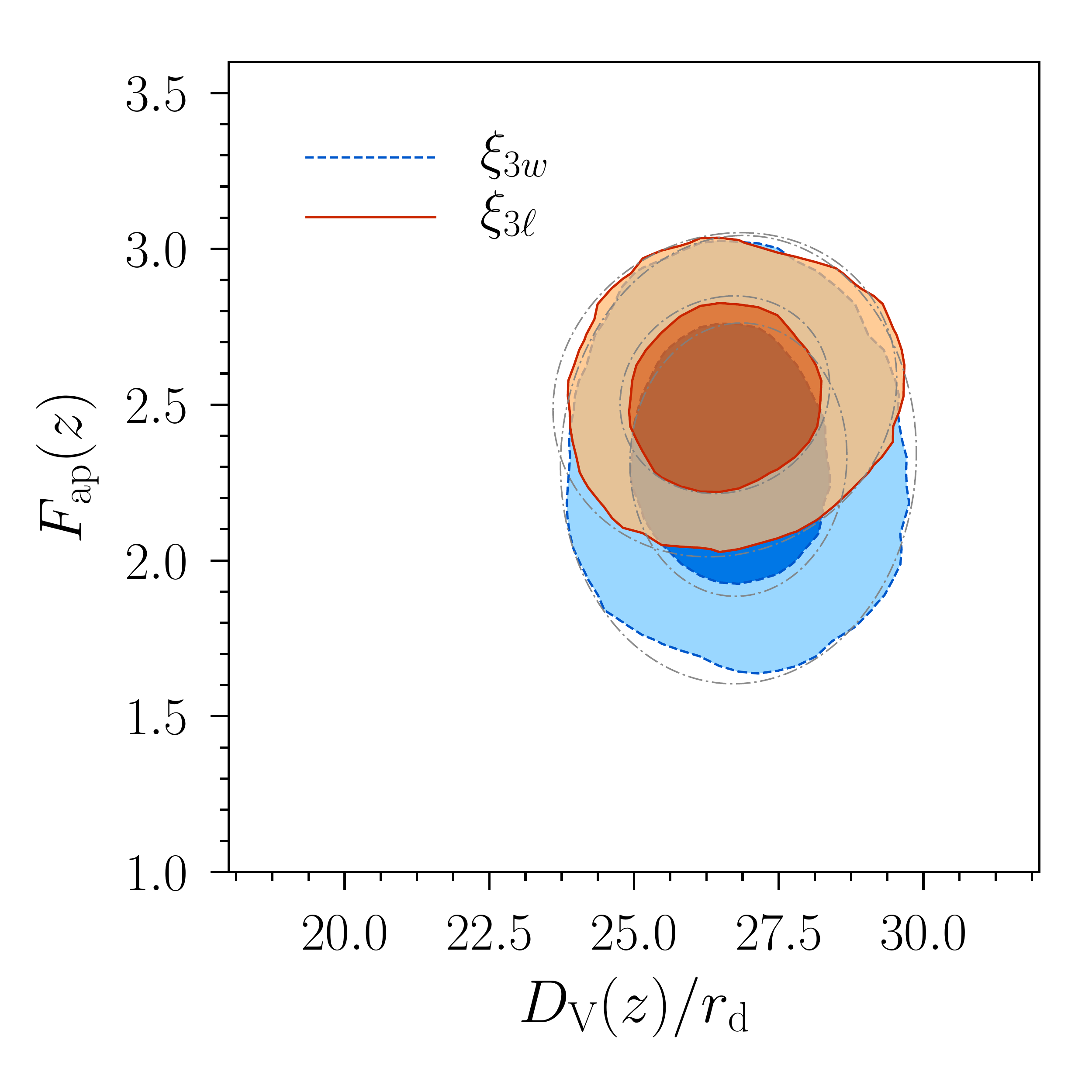}
\includegraphics[width=5cm]{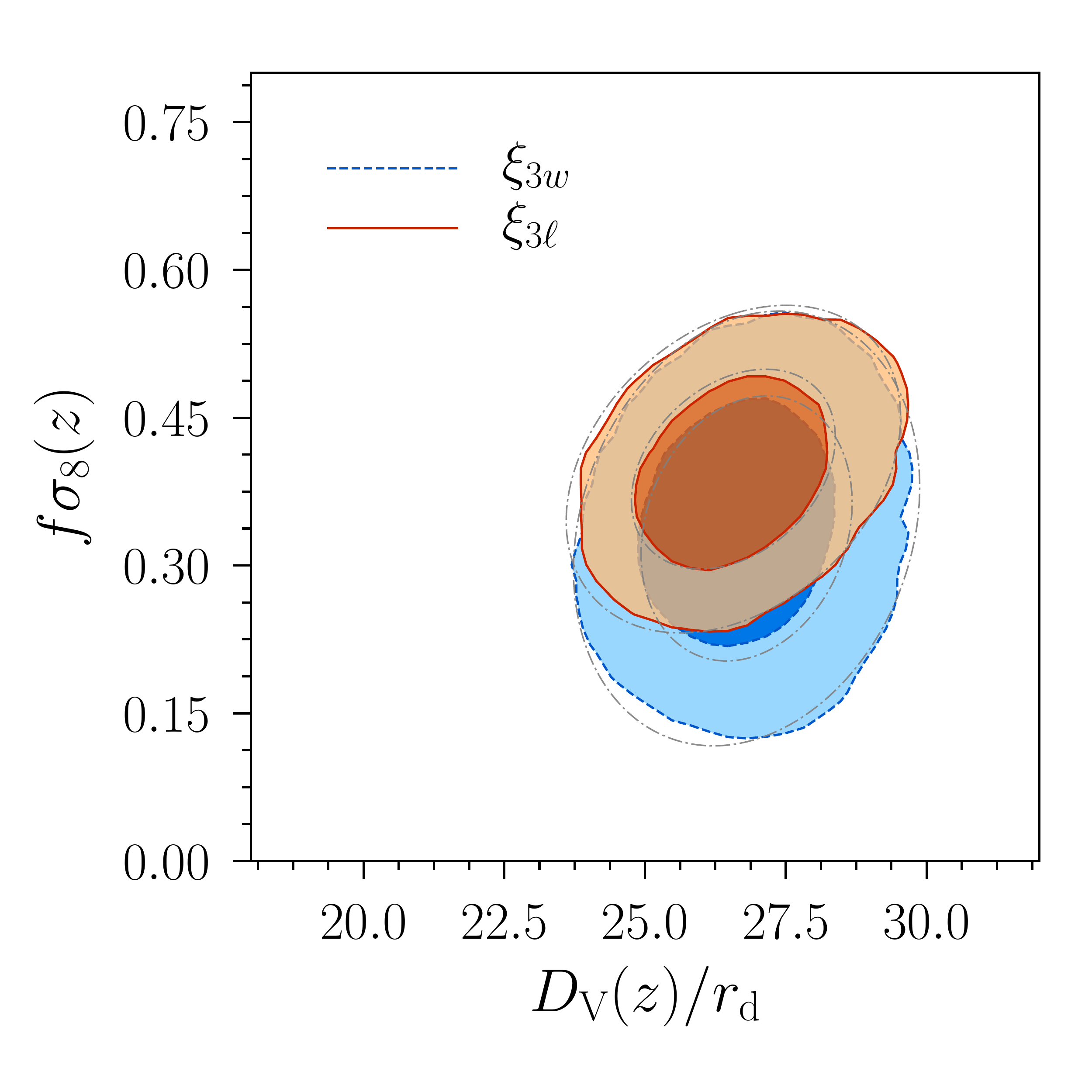}
\includegraphics[width=5cm]{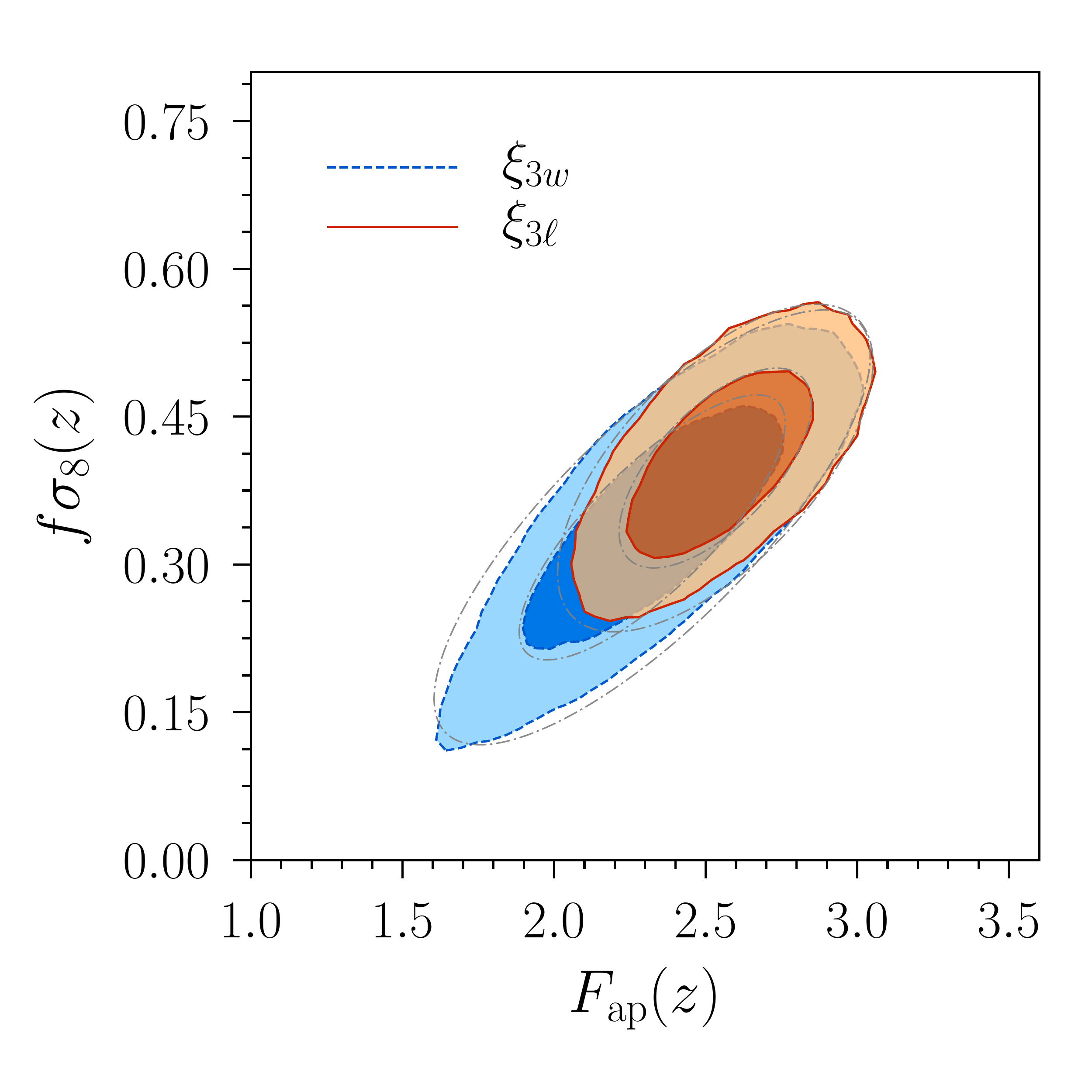}\\
\includegraphics[width=5cm]{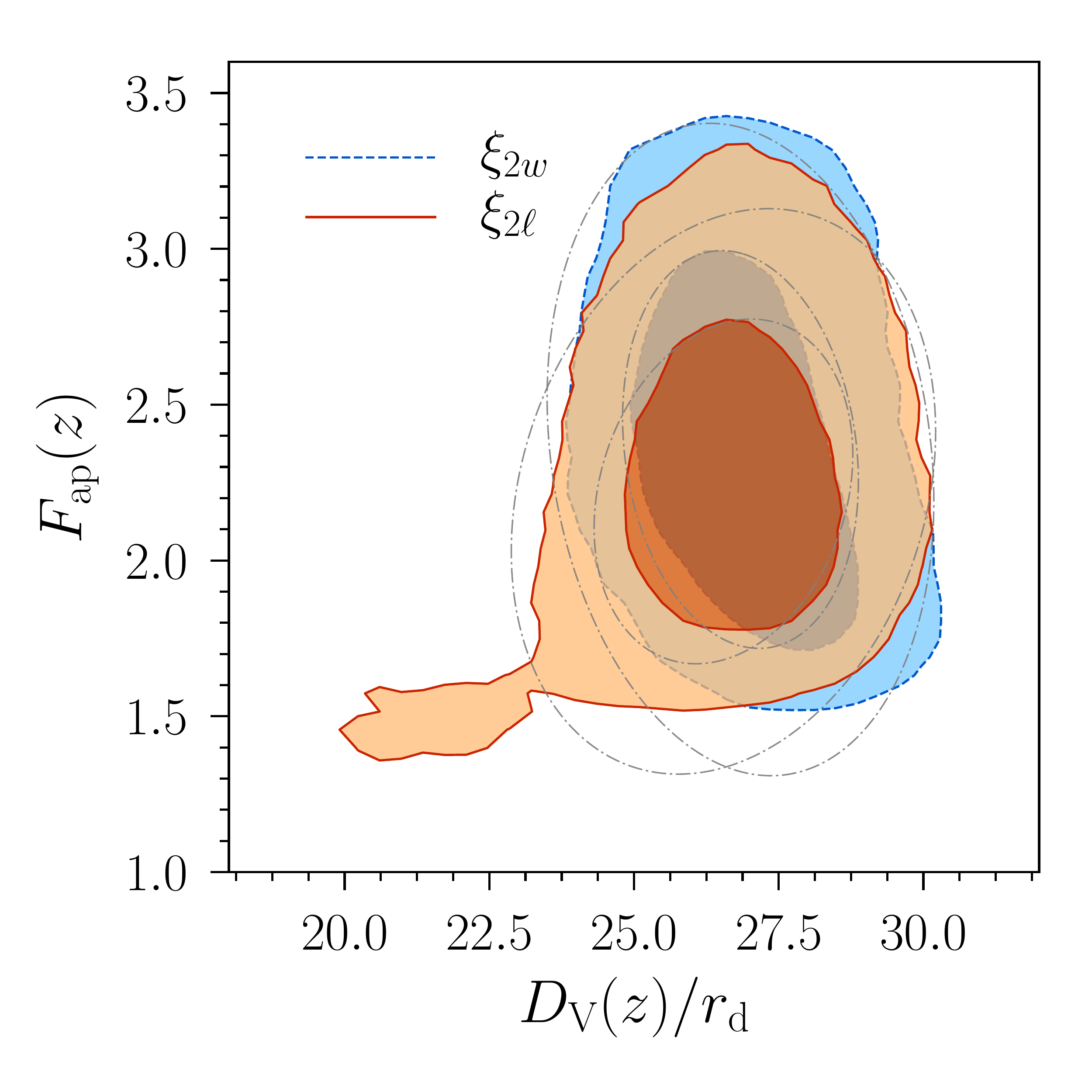}
\includegraphics[width=5cm]{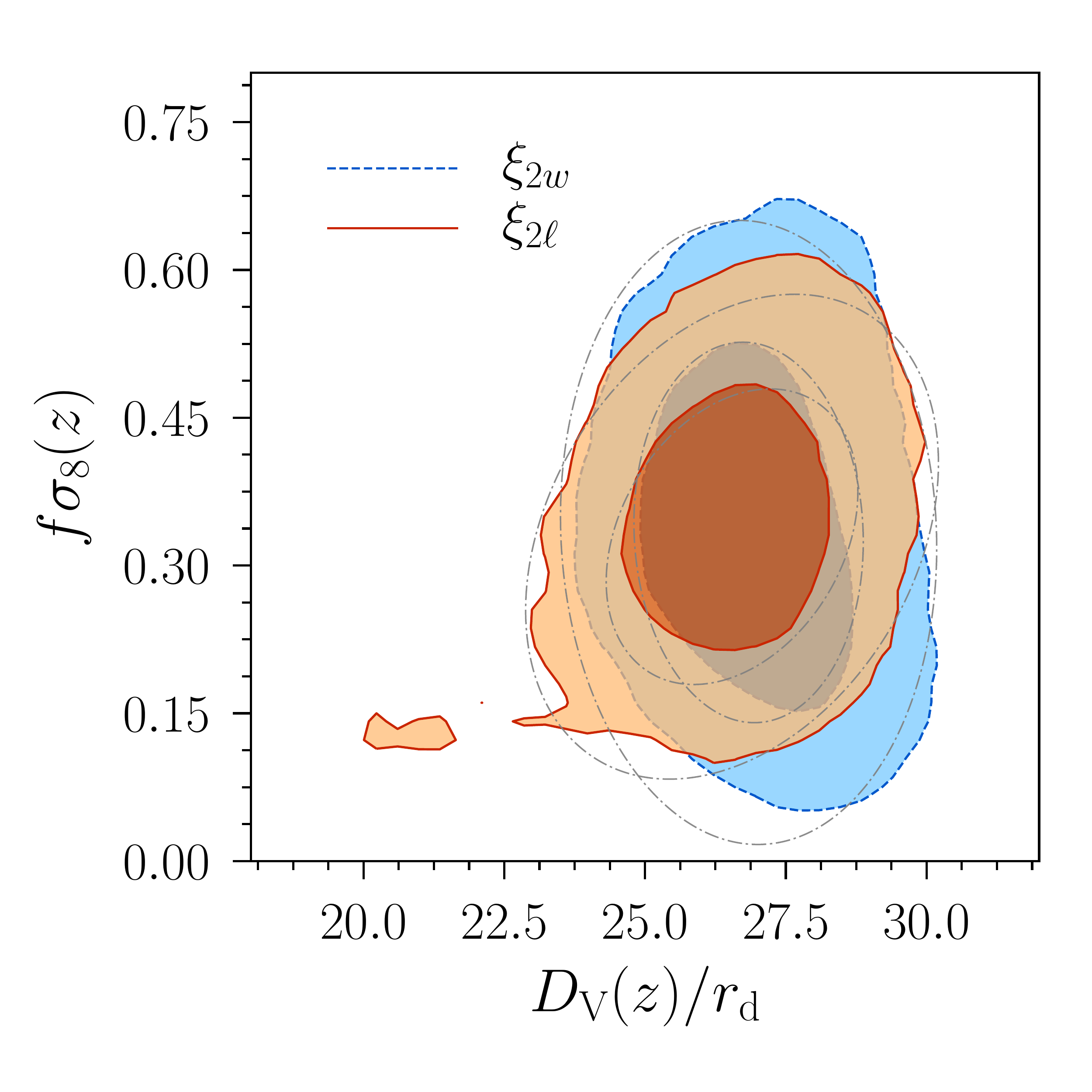}
\includegraphics[width=5cm]{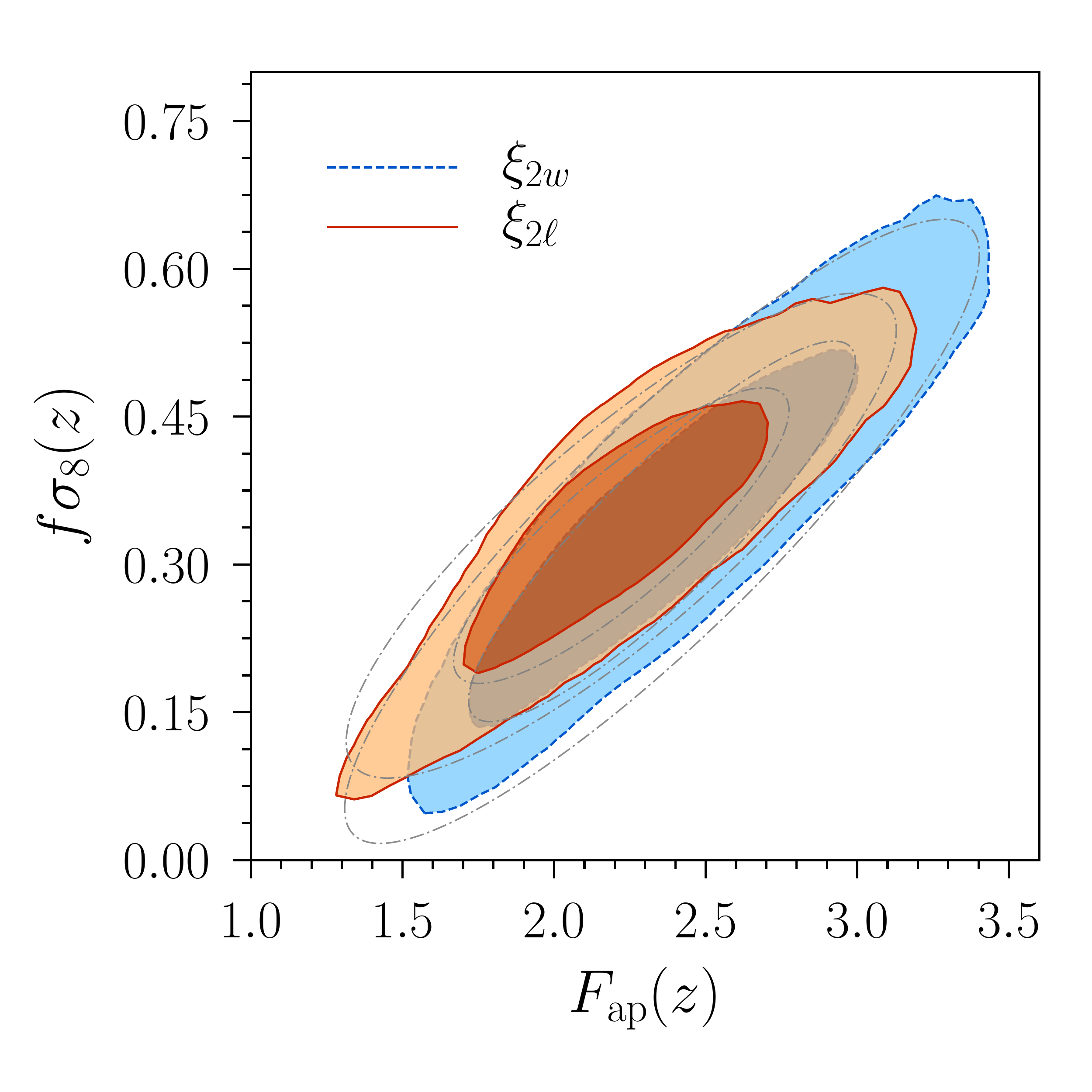}
\caption{Marginalized two-dimensional posterior distributions of the parameters  
$D_{\rm V}/r_{\rm d}(z_{\rm eff})$, $F_{\rm AP}(z_{\rm eff})$, and 
$f\sigma_8(z_{\rm eff})$, evaluated at the mean redshift of the eBOSS quasar sample, 
$z_{\rm eff}=1.52$. The contours represent the $68\%$ (darker regions) and $95\%$ 
(lighter regions) confidence levels. The blue contours show the results obtained 
from measurements of clustering wedges, while the orange contours correspond to those 
recovered from fits to Legendre multipoles.  
The upper panels shows the constraints for three multipoles $\xi_{\ell=0,2,4}(s)$ and 
wedges $\xi_{\rm 3w}(s)$ and the lower panels represent the results obtained by 
fitting $\xi_{\ell=0,2}(s)$ and $\xi_{\rm 2w}(s)$ .
In all cases, the dotted lines represent the Gaussian 
approximation to the full posterior distributions.}
\label{fig: 2d_param_obs}	   
\end{figure*}

\begin{table*}
\centering
\caption{ 
Mean values and 68 $\%$ confidence level (CL) on $D_{\rm V}(z)/r_{\rm d}$, $F_{\rm AP}(z)$ and $f\sigma_8(z)$ recovered from the fits to different clustering statistics measured 
from the eBOSS DR14 quasar LSS sample. }
\begin{tabular}{ | l | l | l | l | l | l |l|}
\hline
Stat. & $\qquad N_{w_i}=3$&   $\qquad N_{\ell_i}=3$&  $\qquad N_{w_i}=2$&  $\qquad N_{\ell_i}=2$&\\
\hline
$D_{\rm V}/r_{\rm d}$ & $26.72\pm 1.13$  & $26.47\pm 1.10$ & $26.72\pm 1.21$&  $26.43\pm 1.19$& \\
$F_{\rm AP}$& $2.332\pm 0.281$&  $2.529\pm0.200$& $2.377\pm 0.433$& $2.233\pm 0.344$& \\
$f\sigma_8$ &$0.339\pm 0.083$&  $0.396\pm 0.063$& $0.339\pm 0.125$&  $0.331\pm 0.092$&\\
\hline
\end{tabular}
\label{tab: RSD_BAO_params}
\end{table*}

Fig. \ref{fig: 2d_param_obs} shows the two-dimensional posterior distributions on different 
combinations of $D_{\rm V}(z_{\rm eff})/r_{\rm d}$, $F_{\rm AP}(z_{\rm eff})$, and 
$f\sigma_8(z_{\rm eff})$ obtained from the eBOSS DR14 quasar sample. 
The blue contours indicate the results inferred from clustering wedges and the orange contours are those obtained from Legendre multipoles. The upper panels present the constraints obtained from the fits to $\xi_{\ell=0,2,4}(s)$ and $\xi_{\rm 3w}(s)$ cases, while the lower panels show the posterior distributions recovered from the monopole-quadrupole pair alone (i.e. excluding information from the hexadecapole) and from two clustering wedges $\xi_{\rm 2w}(s)$. Table~\ref{tab: RSD_BAO_params} lists the one-dimensional marginalized constraints
on $D_{\rm V}/r_{\rm d}$, $F_{\rm AP}$ and $f\sigma_8$ obtained in all cases. 

A comparison of the upper and lower panels of Fig.~\ref{fig: 2d_param_obs} illustrates
the impact that adding the hexadecapole, or using three clustering 
wedges, has on the obtained constraints. The additional information on the full shape 
of $\xi(s,\mu)$ reduces the degeneracy between 
$F_{\rm AP}$ and $f\sigma_8(z_{\rm eff})$, leading to significantly 
tighter results. Fig.~\ref{fig: 2d_param_obs} and Table~\ref{tab: RSD_BAO_params} also show that the fits to three multipoles $\xi_{\ell=0,2,4}(s)$ provide tighter constraints than those obtained using three wedges $\xi_{\rm 3w}(s)$. This result is in agreement with our tests on the EZmocks presented in Section~\ref{subsec: model test}, which also revealed a difference of the same level in the allowed parameter ranges recovered from multipoles and wedges.  

The dotted ellipses in Fig.~\ref{fig: 2d_param_obs} represent the Gaussian approximation 
of the full parameter posterior distributions, based on their corresponding mean values, 
$\bar{\bf D}$, and  covariance matrices, 
$\boldsymbol{\mathrm{\Sigma}}$, as inferred from our MCMC.
Although the results obtained from the measurements of two Legendre 
multipoles or wedges are clearly non-Gaussian, the constraints obtained when 
fitting $\xi_{\ell=0,2,4}(s)$ or $\xi_{\rm 3w}(s)$ are well described by Gaussian 
profiles. This behaviour means that these distributions can be well approximated by
\begin{equation}
\mathcal{P}({\bm \lambda})\propto\exp\left[-\left(\bar{{\bf D}}-{\bf D}_{\rm theo}({\bm \lambda})\right)^t
\boldsymbol{\mathrm{\Sigma}}^{-1}\left(\bar{{\bf D}}-{\bf D}_{\rm theo}({\bm \lambda})\right)\right],
\label{eqn: gauss_post}
\end{equation}
where ${\bf D}_{\rm theo}({\bm \lambda})$ represents the theoretical prediction of the 
distance and growth measurements ${\bf D}$ obtained for the cosmological parameters ${\bm \lambda}$.
As discussed in Section~\ref{subsec: model test}, we treat the constraints derived from the 
fits to the Legendre multipoles $\xi_{\ell=0,2,4}(s)$ as our main parameter constraints.
This information can be compressed in the mean parameter values obtained in this case and their corresponding covariance matrix. However, the resulting distribution would only represent the statistical uncertainties associated with our measurements, without taking into account any potential systematic errors.

We use the results from our fits to the OuterRim HOD mocks to define a systematic error budget 
of our eBOSS measurements. We follow a conservative approach and take the largest deviation between 
our results from the fits to three Legendre Multipoles and their fiducial values as listed in 
Table~\ref{tab: sys_test_OR} and obtain 
$\Delta \alpha_{\perp}= 0.018$, $\Delta \alpha_{\parallel}= 0.036$, $\Delta f\sigma_8= 0.046$.
As in our companion papers, we assume that these systematic errors are independent. 
These values are transformed into the $D_{\rm V}$--$F_{\rm AP}$ basis in which we express 
our results using the Jacobian transformation.

\begin{table*}
\centering
\caption{Parameter covariance matrix for $D_{\rm V}/r_{\rm d}$, $F_{\rm AP}$ and $f\sigma_8$ 
on the BAO and RSD analysis with different statistics configuration. The numbers in the brackets are the systematic error derived based on the test of OuterRim simulation in terms of AP parameters and transformed into $D_{\rm V}$-$F_{\rm AP}$ basis.}	
\begin{tabular}{ | l | l | l | l |  }
\hline
Parameter & $D_{\rm V}/r_{\rm d}$ & $F_{\rm AP}$ & $f\sigma_8$\\
\hline
$D_{\rm V}/r_{\rm d}$ & $1.32508 (+1.80486\cdot 10^{-1})$ &$ 2.03452\cdot 10^{-2} (-1.17239\cdot 10^{-2}) $& $2.35976\cdot 10^{-2}$ \\
$F_{\rm AP}$    &- 	 & $4.05164\cdot 10^{-2} (+7.61549\cdot 10^{-3})$  & $8.40644\cdot 10^{-3}$   \\
$f\sigma_8$    &- & - & $4.12582\cdot 10^{-3} (+2.11600\cdot 10^{-3}) $\\
\hline
\end{tabular}
\label{tab: RSD_BAO_parcov}
\end{table*}
The final covariance matrix representing our constraints from three Legendre multipoles
$\xi_{\ell=0,2,4}(s)$, taking into account both statistical and systematic errors (the numbers in the brackets), 
is listed in Table \ref{tab: RSD_BAO_parcov}.
Our measurements can then be combined with the information from additional data sets by means of a Gaussian likelihood function of the form of equation~(\ref{eqn: gauss_post}), with the  mean parameter values given by the second column of Table~\ref{tab: RSD_BAO_params}, and the covariance matrix given in Table \ref{tab: RSD_BAO_parcov}, which represent the main result of this paper.

\begin{figure*}
\centering
\includegraphics[width=5cm]{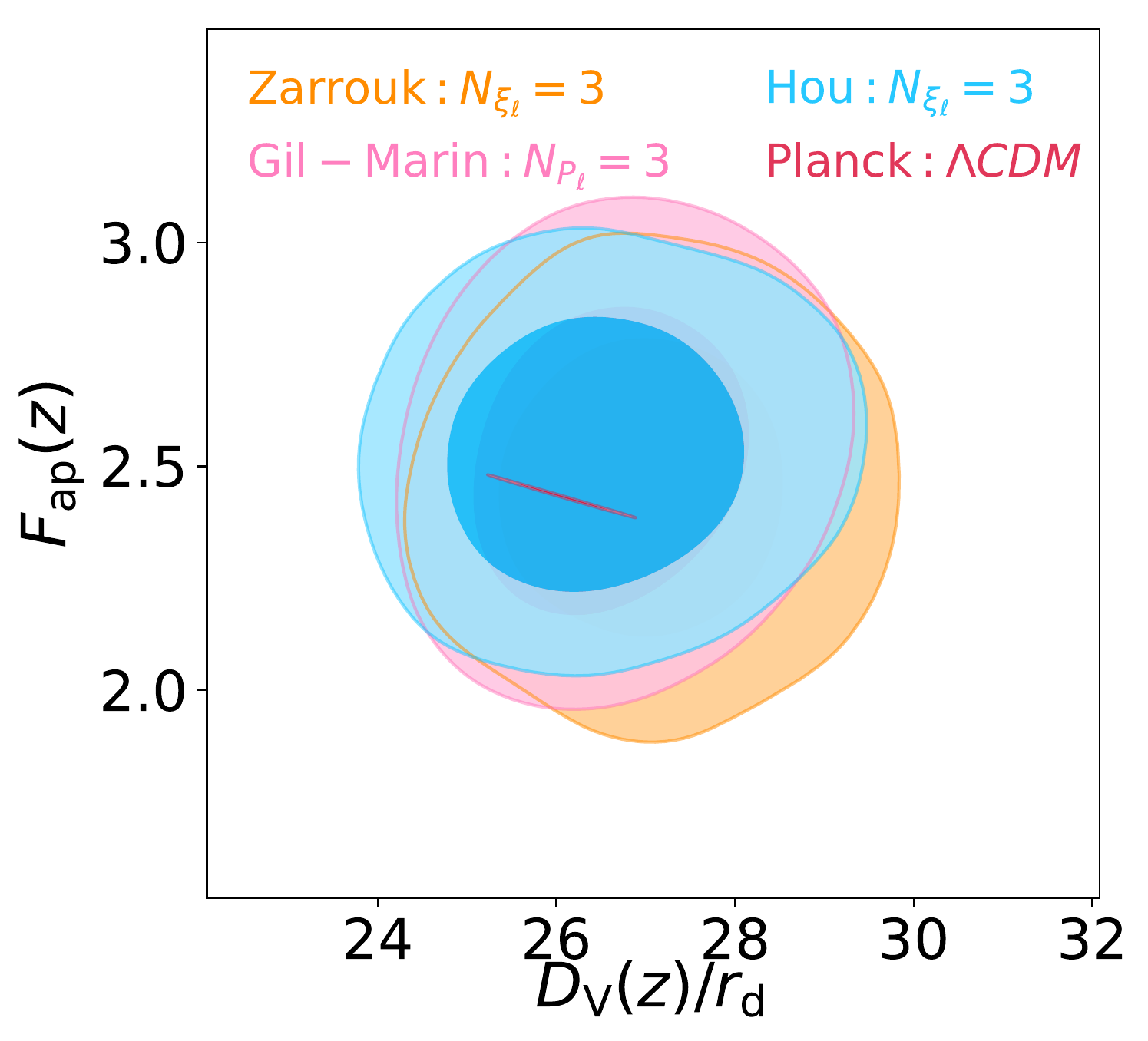}
\includegraphics[width=5cm]{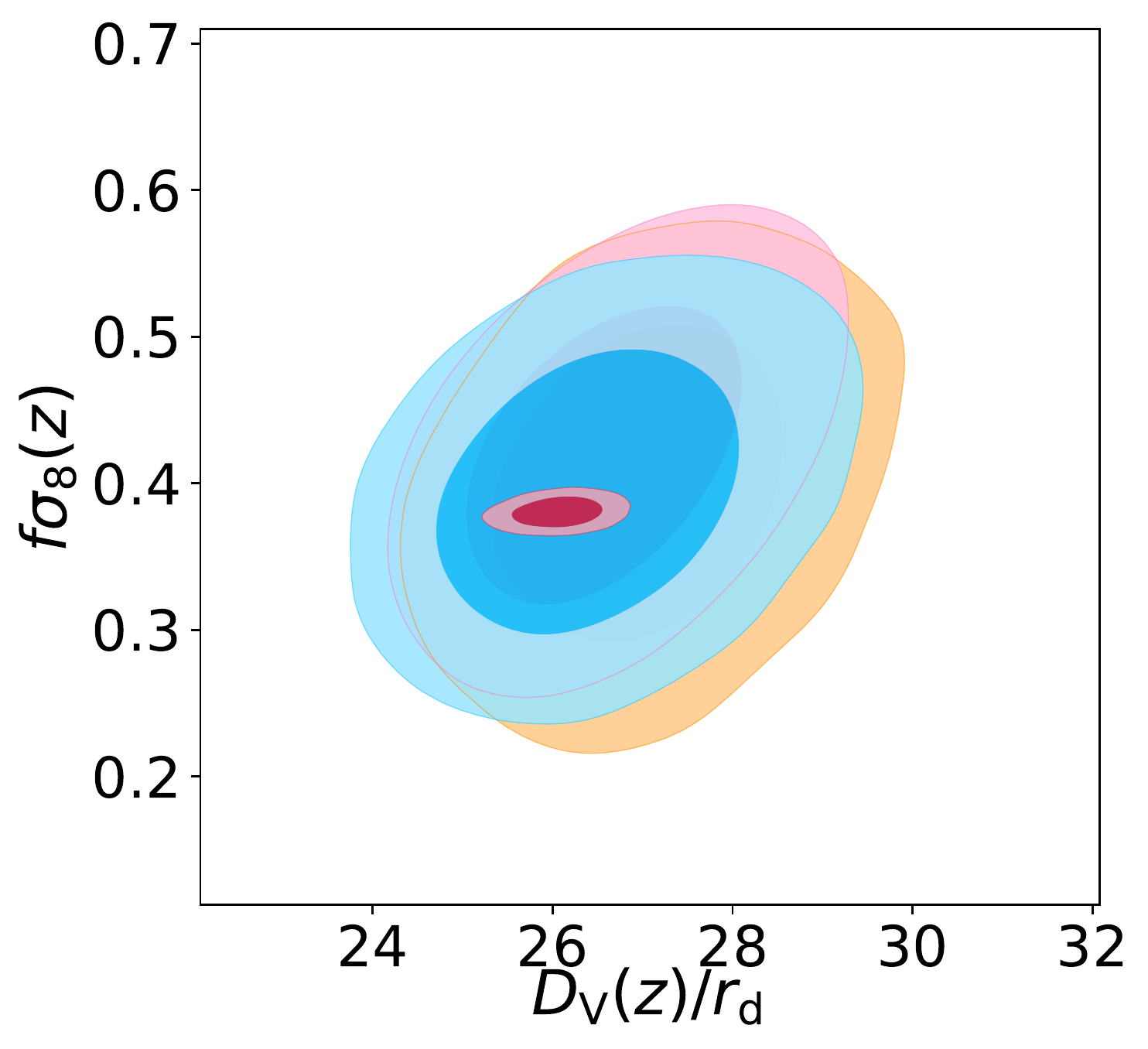}
\includegraphics[width=5cm]{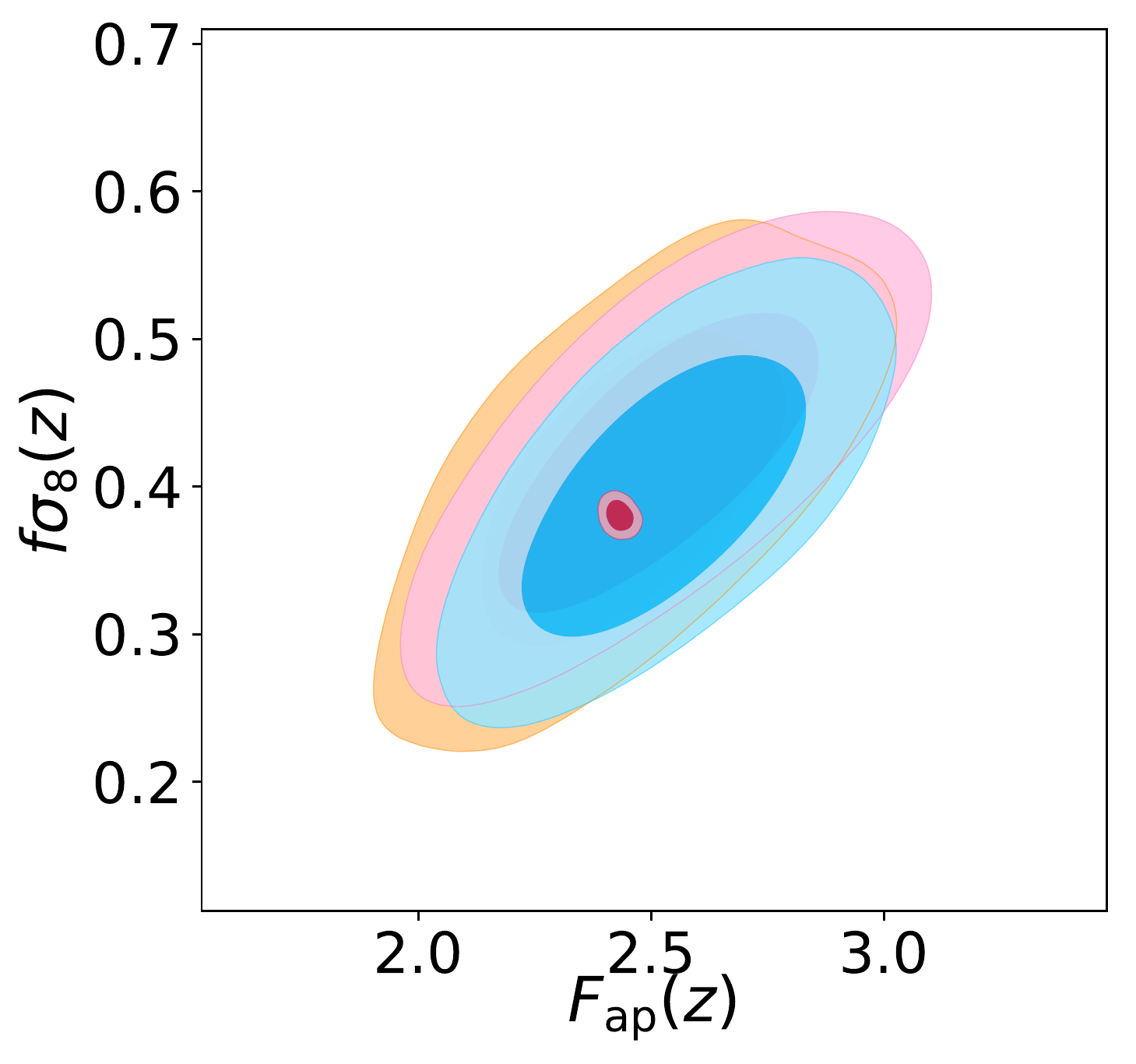}
\caption{Constraints on parameters $f\sigma_8(z_{\rm eff})$, $D_{\rm V}(z_{\rm eff})/r_{\rm d}$ and $F_{\rm Ap}(z_{\rm eff})$ at effective redshift $z_{\rm eff}=1.52$ from different companion papers using the same DR14 LSS quasar dataset. The figure present comparison in terms of 3- Legendre multipoles in both configuration and Fourier space. The blue contour is the result based on the analysis in this paper and the yellow contour is from \citep{Zarrouk-2017}, where both are analysed in configuration. The pink contour is from \citep{Gil-Marin-2017}, analysed in Fourier space. The red contour is from the Planck prediction.}
\label{fig: Consensus}	   
\end{figure*}

\subsection{Comparison with our companion analyses}
\label{subsec: companion}

This work is part of a set of complementary RSD analyses \citep{Zarrouk-2017, Gil-Marin-2017, 
Ruggeri-2017, Zhao-2017}. 
Of these studies, the analyses of \citet{Zarrouk-2017} and \cite{Gil-Marin-2017} are more
closely related to ours. 
\citet{Zarrouk-2017} performed an analysis of the full shape of the configuration-space
Legendre multipoles and clustering wedges for scales between 
16 and $138\,h^{-1}{\rm Mpc}$ using a model based on convolution Lagrangian perturbation 
theory \citep{Carlson-2013, Wang-2014} and the Gaussian streaming model 
\citep{Peebles-1980,Fisher-1995,Scoccimarro-2004, Reid-2011}. 
\citet{Gil-Marin-2017} applied a model based on \citet*{Taruya-2010} to the Legendre multipoles
in Fourier-space, $P_{\ell}(k)$, for $\ell=0,2,4$ up to scales of $k = 0.3\,h{\rm Mpc}^{-1}$. 
{
These methods represent the results at one effective redshift bin and hereafter we refer as the conventional analyses.
}
We focus here on a comparison among the conventional analyses.

Fig.~\ref{fig: Consensus} presents a comparison of the two-dimensional posterior 
distributions of $D_{\rm v}(z_{\rm eff})/r_{\rm d}$, $ F_{\rm AP}(z_{\rm eff})$, and 
$f\sigma_8(z_{\rm eff})$ at $z_{\rm eff}=1.52$ from \citet{Zarrouk-2017} and 
\citet{Gil-Marin-2017} and our results based on the Legendre
multipoles $\xi_{\ell}(s)$, with $\ell=0,2,4$ for $16\,h^{-1}{\rm Mpc}<s<160\,h^{-1}{\rm Mpc}$.
Despite the differences in the range of scales and data used, as well as on the modelling 
of non-linear evolution, bias and RSD implemented in these analyses, the derived constraints are in excellent agreement with each other, demonstrating the robustness of the results.
The red contours in the same figure represent the constraints inferred from 
the Planck CMB measurements under the assumption of a flat $\Lambda$CDM cosmology. 
The CMB constraints, which are strongly model-dependent, are in good agreement with the 
results inferred from the clustering analyses of the eBOSS LSS quasar sample, demonstrating the 
consistency between these datasets within the context of the $\Lambda$CDM model.

{ In additional, \cite{Gil-Marin-2017} have performed test by splitting the sample into three redshift bins. They found the result is not significantly affected either using a single bin or three bins, which indicates that representing the given sample one effective redshift is valid.}

{ Complementing these conventional RSD analyses, \citet{Ruggeri-2017} and \cite{Zhao-2017} applied a redshift-dependent weighting scheme to the Legendre multipoles of the power spectrum to compress the information along the redshift direction. 
 A more detailed comparison between the results of all companion
papers, including those implementing redshift weighting schemes can 
be found in \citet{Zhao-2017, Zarrouk-2017}. The consistency between the conventional analysis and the redshift-weighted method also shows that representing the sample at the effective redshift does not introduce significant systematic errors.}

\section{Conclusions}
\label{sec: conclusion}

We have presented an analysis of the anisotropic clustering of DR14 eBOSS quasar sample in 
configuration space. 
Using quasars as tracers of the LSS has the advantage that it allows one to extend clustering analyses to higher redshift than using galaxies. 
We projected the information of the full two-dimensional correlation function $\xi(s,\mu)$
of the eBOSS quasar sample into Legendre multipoles $\xi_{\ell}(s)$ with $\ell=0,2,4$ and 
clustering wedges measured using two and three $\mu$-bins, $\xi_{\rm 2w}(s)$ and 
$\xi_{\rm 3w}(s)$.

Our study makes use of a state-of-the-art model of non-linear evolution, bias, and RSD 
that was previously applied to the analysis of the final BOSS galaxy samples 
\citep{Sanchez-2017a, Grieb-2017a,Salazar-Albornoz-2017}, modified to account for 
non-negligible redshift errors.
When comparing these theoretical predictions against the measurements of the Legendre
multipoles and clustering wedges of the eBOSS sample we use the likelihood function of 
\citet{Sellentin-2016}. This recipe correctly accounts for the noise in our estimates of 
the covariance matrices, which were derived from a set of $1\,000$ synthetic eBOSS quasar 
catalogues. 
The tests of our analysis methodology on these mocks catalogues show that it can 
extract robust distance and growth of structure measurements from our eBOSS 
quasar clustering measurements for scales $s \gtrsim 20\,h^{-1}{\rm Mpc}$. 

We also test our model using a full N-body simulation and define the systematic error based on the test result. Adding the systematic error inflates the error budget on $f\sigma_8$ by about $25\%$. Future investigation from both sides of the simulation and modelling will help to decrease this error.

Our tests demonstrate that the analysis of the first three non-zero Legendre multipoles provides tighter constraints than the other statistics we considered. For this reason, we define the constraints derived from $\xi_{\ell=0,2,4}(s)$ as the main result of our analysis.
These constraints can be expressed as measurements of the parameter combinations 
$D_{\rm V}(z_{\rm eff})/r_{\rm d}$, $F_{\rm AP}(z_{\rm eff})$ and $f\sigma_8(z_{\rm eff})$ at the effective redshift of the eBOSS LSS quasar sample, $z_{\rm eff}=1.52$. 
The posterior distribution of these parameters is well described by a Gaussian and can be correctly represented by the mean values of these parameters and their corresponding covariance matrix, which we provide here.

Our analysis is part of a set of papers focused on extracting geometric and growth of
structure constraints from the eBOSS quasar sample \citep{Zarrouk-2017, Gil-Marin-2017, 
Ruggeri-2017, Zhao-2017}. 
In particular, the analyses of \citet{Gil-Marin-2017} and \citet{Zarrouk-2017}, 
who considered the information of two-point clustering measurements in Fourier and
configuration space obtained from the full redshift range $0.8 < z< 2.2$, are the ones most similar to our study. A comparison of our results with 
those of the companion papers shows remarkable consistency, demonstrating the robustness of 
the obtained results with respect to choice of data and the details of modelling 
implemented.

The results from our analysis and those of our companion papers demonstrate that 
quasars can be used as robust tracers of the large-scale clustering pattern. 
The methodologies previously used to extract cosmological information from 
anisotropic clustering measurements based on galaxy samples are
applicable to quasars as well, providing a powerful cosmological probe at high 
redshift.
The application of these techniques to future quasar samples from eBOSS and other surveys, 
which will cover larger volumes, will provide a more complete view of the expansion and growth of structure histories of our Universe.

\section*{Acknowledgements}

JH and AGS thank Mart\'{i}n Crocce for the contribution to the development of the model used in this analysis.
JH and AGS acknowledge Daniel Farrow, Fabrizio Finozzi, Martha Lippich and Francesco Montesano for the useful discussion. JH thanks Hao Ding for the support. 
JH and AGS acknowledge support from the Trans-regional Collaborative Research 
Centre TR33 `The Dark Universe' of the German Research Foundation (DFG).

\noindent G.R. acknowledges support from the National Research Foundation of Korea (NRF) through Grant No. 2017077508 funded by the Korean Ministry of Education, Science and Technology (MoEST), and from the faculty research fund of Sejong University in 2018.

\noindent GBZ is supported by NSFC Grants 1171001024 and 11673025. GBZ is also supported by a Royal Society Newton Advanced Fellowship, hosted by University of Portsmouth.

\noindent Funding for SDSS-IV has been provided by the
Alfred P. Sloan Foundation and Participating Institutions. SDSS is managed by the Astrophysical Research Consortium for the Participating Institutions of the SDSS Collaboration including the Brazilian Participation Group, the Carnegie Institution for Science, Carnegie Mellon University, the Chilean Participation Group, the French Participation Group, Harvard-Smithsonian Center for Astrophysics, Instituto de Astrof\'{i}sica de Canarias, The Johns Hopkins University, Kavli Institute for the Physics and Mathematics of the Universe (IPMU) / University of Tokyo, Lawrence Berkeley National Laboratory, Leibniz Institut f\"{u}r Astrophysik Potsdam (AIP), Max-Planck-Institut für Astronomie (MPIA Heidelberg), Max-Planck-Institut f\"{u}r Astrophysik (MPA Garching), Max-Planck-Institut f\"{u}r Extraterrestrische Physik (MPE), National Astronomical Observatories of China, New Mexico State University, New York University, University of Notre Dame, Observat\'{o}rio Nacional / MCTI, The Ohio State University, Pennsylvania State University, Shanghai Astronomical Observatory, United Kingdom Participation Group, Universidad Nacional Aut\'{o}noma de M\'{e}xico, University of Arizona, University of Colorado Boulder, University of Oxford, University of Portsmouth, University of Utah, University of Virginia, University of Washington, University of Wisconsin, Vanderbilt University, and Yale University.

\noindent This research used resources of the National Energy Research Scientific Computing Center, a DOE Office of Science User Facility supported by the Office of Science of the U.S. Department of Energy under Contract No. DE-AC02-05CH11231.

\bibliographystyle{mnras}

\bibliography{reference_paper}

\bsp	

\appendix

\section{Likelihood profile and uncertainty correction}
\label{appendix: likelihood}
As discussed in sec. \ref{sec: likelihood}, the matrix inverse operation the covariance 
covariance matrix can lead to a non-Gaussian likelihood profile if the covariance matrix is estimated 
from a limited number of mocks. The modified likelihood profile asymptotically approaches 
the simple Gaussian recipe as the number of mocks increases. 
If a Gaussian likelihood profile is assumed, the noise due to the limited number of mocks 
must be propagated into the final parameter constraints. In this case, the obtained parameter
covariance matrix needs to be rescaled by a factor \citep{Percival-2014},
\begin{equation}
M=\frac{1+B(N_{\rm b}-N_{\rm p})}{1+A+B(N_{\rm p}+1)},
\label{eqn: M_factor}
\end{equation}
where,
\begin{align}
A=\frac{2}{(N_{\rm m}-N_{\rm b}-1)(N_{\rm m}-N_{\rm b}-4)}\\
B=\frac{(N_{\rm m}-N_{\rm b}-2)}{(N_{\rm m}-N_{\rm b}-1)(N_{\rm m}-N_{\rm b}-4)}.
\label{eqn: AB_factor}
\end{align}
with $N_{\rm b}$ being the number of bins in the data vector, $N_{\rm p}$ being the number of free parameters, and $N_{\rm m}$ being the number of simulations used to estimate 
the covariance matrix.
Table \ref{tab: model_test_EZ} lists the correction factors $M$ corresponding to the 
Legendre multipoles and clustering wedges for different rage of scales ranges. 

\begin{table}
\centering
\caption{ Factors to correct the parameter covariance matrix when different scales are included 
in the analysis. The values of the minimum scales are expressed in $h^{-1} {\rm Mpc}$. In all 
cases, the maximum scale considered was $s_{\rm max}=160\,h^{-1} {\rm Mpc}$, the covariance
matrix were estimated using $N_{\rm m}= 1\,000$ mock catalogues, and the fits
included $N_{\rm p}=7$ free parameters }	
\begin{tabular}{ | c | c | c| }
\hline
 $s_{\rm min}$ &  $N_{\rm b}$  & $M$ \\
\hline
  $8$   &$57$ & $1.0219$\\
  $16$  &$54$ & $1.0203$\\
 $24$  &$51$ & $1.0187$\\
 $32$  &$48$ & $1.0171$\\
\hline
\end{tabular}
\label{tab: model_test_EZ}
\end{table}

\begin{figure}
\centering
\includegraphics[width=8cm]{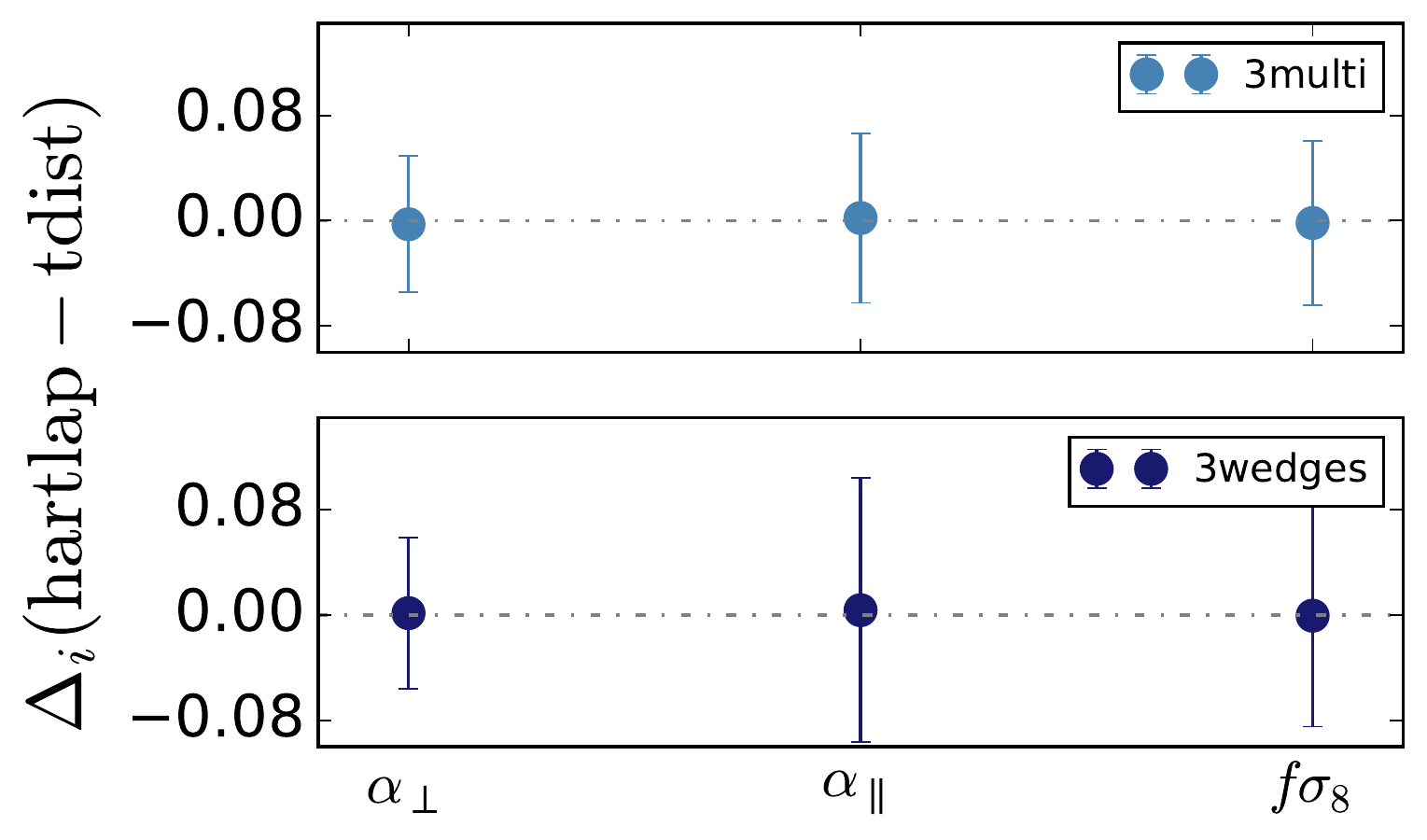}
\caption{Difference on the inferred parameters $\alpha_{\perp}$, $\alpha_{\parallel}$ and $f\sigma_8$ by assuming the likelihood profile for Gaussian$+$Hartlap and modified t-distribution. The errorbar is the statistical error from the marginalized 1d distribution using square-wise sum of both Gaussian and modified t-distribution.}
\label{fig: likelihood_profile}	   
\end{figure}

\begin{figure}
\centering
\includegraphics[width=8cm]{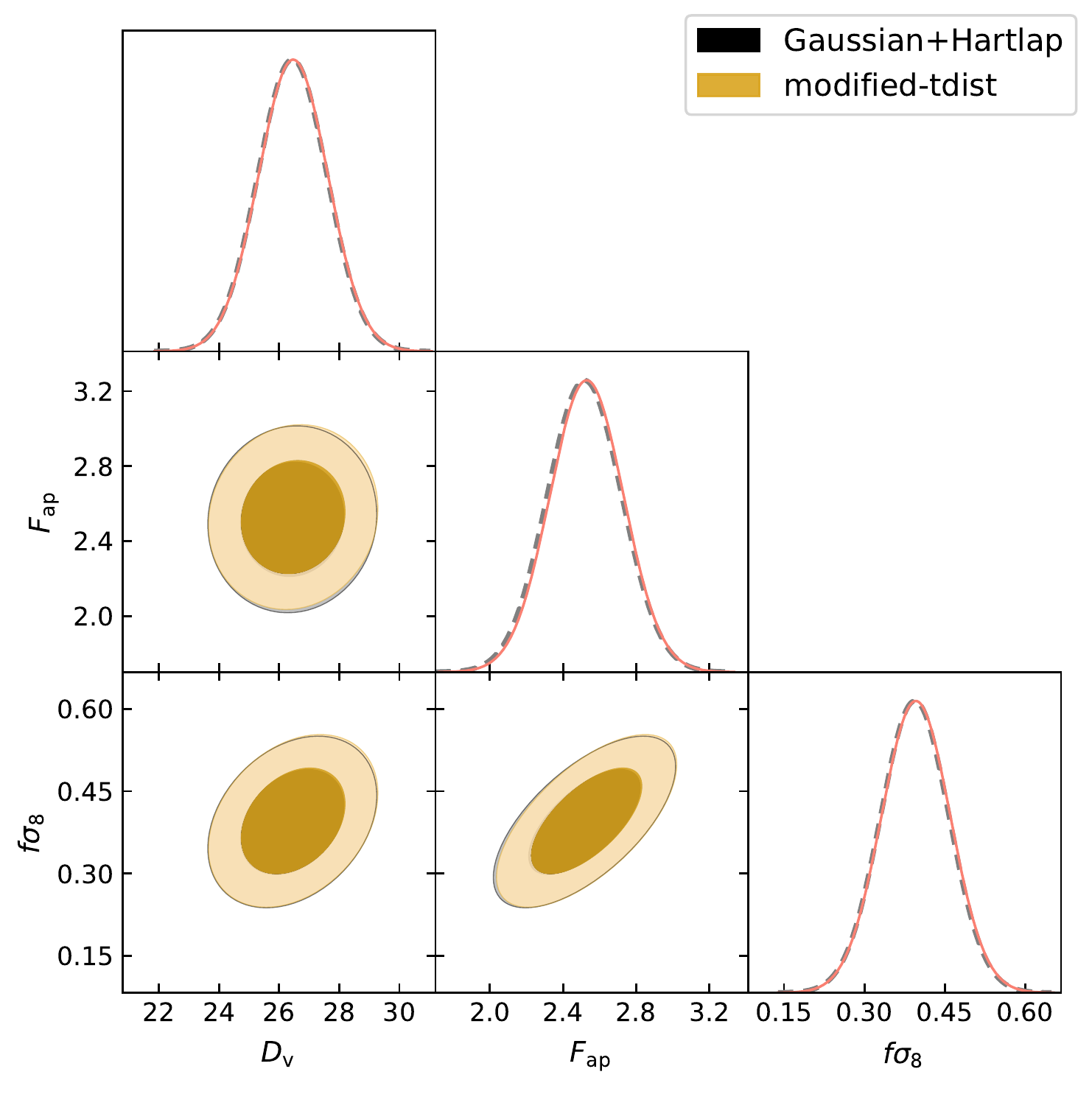}
\caption{Parameter covariance of $f\sigma_8$, $D_{\rm V}$ and $F_{\rm AP}$ using Gaussian likelihood profile with Hartlap correction (black) and modified t-distribution(orange).}
\label{fig: likelihood_profile_2d}	   
\end{figure}

\citet{Sellentin-2016} suggested a modified t-distributed likelihood to account for this effect. 
Here we perform the test on comparing the results obtained from the real eBOSS data using the 
two likelihood profiles, where the covariance matrix is rescaled by the factor of equation 
(\ref{eqn: hartlap}) and the resulting parameter covariance is rescaled by the factor $M$ of 
equation (\ref{eqn: M_factor}). Fig. \ref{fig: likelihood_profile} shows the difference in the AP-parameters and growth rate parameter for Legendre multipoles (upper panel, lighter blue) and clustering wedges (lower panel, darker blue). The errorbars are the statistical error calculated from marginalized 1d distribution by a square-wise sum of both Gaussian and modified t-distribution. Fig. \ref{fig: likelihood_profile_2d} is a direct comparison for the parameter covariance on $f\sigma_8$, $D_{\rm V}$ and $F_{\rm AP}$. There is only a marginal shift in the center of the mean value with $\Delta x$ less than $ 3\% $ of $\sigma$ and the uncertainties on the inferred parameters are comparable with each other.
The agreement between the parameters estimated from the two likelihood profiles confirms that the number of mocks used to estimate the covariance matrix is sufficient for the LSS quasar analysis. 

\label{lastpage}
\end{document}